\documentclass[trackchanges, twocolumn]{aastex701}
\usepackage{appendix}
\usepackage{amsmath,amsfonts,amssymb}
\usepackage{hyperref}
\usepackage[percent]{overpic}
\usepackage{xcolor}
\usepackage{booktabs}
\usepackage{capt-of}
\usepackage{multirow}
\usepackage{comment}
\begin{document}

\title{A Four-dimensional Model-agnostic Probe into the Astrophysical Origins of Binary Black Hole Subpopulations}

\author[0000-0002-7322-4748]{Anarya\, Ray}
\affiliation{Center for Interdisciplinary Exploration and Research in Astrophysics, Northwestern University, 1800 Sherman Avenue, Evanston, IL 60201, USA}
\affiliation{NSF-Simons AI Institute for the Sky (SkAI), 172 E. Chestnut Street, Chicago, IL 60611, USA}
\email[show]{anarya.ray@northwestern.edu}

\author[0000-0001-9236-5469]{Vicky\, Kalogera}
\affiliation{Center for Interdisciplinary Exploration and Research in Astrophysics, Northwestern University, 1800 Sherman Avenue, Evanston, IL 60201, USA}
\affiliation{NSF-Simons AI Institute for the Sky (SkAI), 172 E. Chestnut Street, Chicago, IL 60611, USA}
\affiliation{Department of Physics and Astronomy, Northwestern University, 2145 Sheridan Road, Evanston, IL 60208, USA}
\email{vicky@northwestern.edu}
\begin{abstract}

There is strong evidence of multiple binary black hole~(BBH) subpopulations in the cumulative gravitational wave catalog by the LIGO-Virgo-KAGRA collaboration that likely originate through distinct evolutionary channels. The astrophysical interpretation of this complex underlying population is subject to theoretical uncertainties in treatments of binary stellar evolution, core collapse, and host environments. Due to a lack of robust predictions and the sheer diversity of plausible features, strongly modelled population analyses often lead to prior-driven conclusions. On the other hand, flexible alternatives are often difficult to scale in higher dimensions, which can lead to a loss of critical information on astrophysically meaningful correlations. In this \textit{Letter}, we present the first data-driven reconstruction of the joint four-dimensional distribution of BBH primary masses, mass ratios, effective inspiral and effective precessing spin parameters, which yields novel model-agnostic constraints on the astrophysical origins of BBH subpopulations. We characterize four distinct subpopulations spanning different ranges of BBH masses and report new correlations in these specific mass ranges that are beyond the reach of strongly modeled parametrizations and lower-dimensional data-driven frameworks. Our results unveil novel insights into the abundances of specific subchannels of isolated binary evolution, dynamical assembly, and hierarchical mergers across various mass ranges in the astrophysical BBH population.

%Based on theoretical predictions of population synthesis simulations,

\end{abstract}

%\keywords{\uat{High Energy astrophysics}{739} --- \uat{Stellar astronomy}{1583}}

\section{Introduction} 

The 5th gravitational wave transient catalog~\citep[GWTC-5,][]{LIGOScientific:2026wfs} of the LIGO-Virgo-KAGRA~\citep[LVK,][]{LIGOScientific:2014pky,VIRGO:2014yos, KAGRA:2020agh} detector network has unveiled a complex underlying population of binary black hole~(BBH) mergers~\citep{LIGOScientific:2026ctl}. With nearly 260 confident detections, population analysis has identified multiple clusters in the observed ensemble that likely correspond to specific evolutionary pathways~\citep{Banagiri:2025dmy, Ray:2025xti, Farah:2026jlc, Vijaykumar:2026zjy, Sridhar:2025kvi, Wang:2025nhf, Afroz:2024fzp, Afroz:2025ikg, Galaudage:2026opk, Ray:2026uur, Flanagan:2026ayy, Flanagan:2026btu, Cheng:2026bpc, Alvarez-Lopez:2026ymo, Li:2025iux, Rinaldi:2026nyb, Zevin:2020gbd, Cheng:2023ddt, Colloms:2025hib, Antonini:2024het, Antonini:2025ilj, Tong:2025wpz, Li:2026iae, Zeeshan:2026mpr, Guttman:2026cnv, Padhyegurjar:2026scg, Padhyegurjar:2026slt}. However, theoretical uncertainties in binary stellar evolution, BH formation through stellar collapse, and host environments often lead to a scarcity of robust predictions, limiting the scope of self-consistent astrophysical interpretation of the observed population~\citep[see, e.g.][for reviews]{Mandel:2021smh, Breivik:2025edm, Mandel:2018hfr, Mapelli:2021taw}. 

BBH formation scenarios can be broadly classified into isolated binary evolution in galactic fields and dynamically interacting binaries in dense stellar environments, triple systems, or disks of active galactic nuclei~\citep[see, e.g.,][for reviews]{Mandel:2018hfr, Mapelli:2021taw, Mandel:2021smh}. Dynamically formed binaries can be further grouped into first-generation ``1G+1G" mergers that only comprise BHs formed directly through stellar collapse, and hierarchical mergers that comprise remnants of previous mergers retained by the host environment~\citep[see, e.g.,][for a review]{Gerosa:2021mno}. To disentangle these astrophysical channels and constrain underlying uncertainties, it is necessary to model a large diversity of predicted features in the observed population of BBH masses, redshifts, component spin magnitudes and spin orientations relative to the orbit. Note that distinct astrophysical trends in the distribution of component spins will also manifest in that of the effective aligned and effective precessing spin parameters of BBHs, which are often measured better than component spins for individual detections.

Astrophysical analysis of the BBH population is traditionally approached through two complementary frameworks. Data-driven methods present a model-agnostic characterization of the BBH population, revealing previously unmodeled physics while being less susceptible to prior-driven conclusions and biases~\citep[e.g.,][]{Ray:2023upk, Ray:2024hos, Sridhar:2025kvi, Heinzel:2024, Heinzel:2024hva, Veske:2021, Rinaldi:2021FIGARO, Rinaldi:2023Evolution, Rinaldi:2026nyb, Edelman:2021Mountain, Edelman:2022Cover, Sadiq:2021AdaptiveKDE, Sadiq:2023BinaryVision, LIGOScientific:2025GWTC4Population, Tiwari:2020otp, Tiwari:2021Features}. However, owing to the high complexity of flexible population models, these approaches are difficult to scale in higher dimensions, leading to a loss of information on astrophysically meaningful correlations and biases in the inferred marginal distributions~\citep{Alvarez-Lopez:2025ltt}. Furthermore, they often suffer from large measurement uncertainties in regions of sparse data or in the spaces of weakly measured BBH parameters, which can limit the robustness of interpretation. On the other hand, guided by these inferred trends in the data, targeted parametrizations of channel-specific population properties (that scale easily to higher dimensions) can yield stronger constraints and a coherent astrophysical interpretation of the growing sample of BBH detections~\citep{Ray:2026uur}.

In previous works, we relied on these complementary probes to arrive at a self-consistent astrophysical interpretation of the observed population of BBHs in the previous catalog, GWTC-4. In \cite{Sridhar:2025kvi}, we relied on a three-dimensional data-driven characterization of BBH subpopulations in the space of primary masses, effective inspiral spins, and mass ratios to identify unique astrophysical trends free of prior restrictions. Guided by these findings, we parametrized the BBH population with a three-component mixture model which generalizes to redshifts and effective precessing spins to disentangle the contributions of specific formation channels~\citep{Ray:2026uur}. Our analysis revealed that BBHs in GWTC-4 comprise three subpopulations that correspond to distinct mass-ratio, spin-magnitude, and spin-orientation properties which are consistent with specific formation scenarios. We showed that the observed BBH population likely arises from specific relative abundances of isolated binary evolution, dynamically interacting first-generation BHs (likely including binaries in triple systems), and hierarchical mergers, that vary as a function of mass and evolve over cosmic time~\citep{Ray:2026uur}.

However, these investigations comprise nascent steps towards a fully coherent, robust, and well-established understanding of the origins of the astrophysical BBH population. With the advent of the latest catalog GWTC-5, there is evidence of additional substructure and correlations in the BBH population~\citep{Flanagan:2026btu, Cheng:2026bpc, Alvarez-Lopez:2026ymo, Li:2025iux, Rinaldi:2026nyb, Tiwari:2026bvx} beyond those identified by the three-component parametrizations. This likely indicates different sub-channels of the isolated binary evolution and dynamical assembly contributing to specific subpopulations that were previously attributed to one broad scenario or the other. Furthermore, the exact nature of these multi-dimensional features and correlations, when inferred with strongly modeled parametrizations, can vary significantly depending on the population prior, often leading to contradictory astrophysical interpretations~\citep[see, e.g.,][]{Flanagan:2026btu, Cheng:2026bpc, Alvarez-Lopez:2026ymo} and limited model comparison among theoretical predictions. 

Hence, as the natural next step, a generic data-driven framework for reconstructing the multi-dimensional distribution of BBH masses, effective inspiral and effective-precessing spin parameters, and possibly BBH redshifts, is crucial for corroborating strongly modeled findings and identifying previously unmodeled astrophysical imprints. However, due to the curse of dimensionality, which often leads to computational intractability in flexible multi-dimensional population inference, fully data-driven approaches beyond three dimensions have not been implemented before.

%In this \textit{Letter}, using a non-parametric model for the joint distribution of BBH primary masses, mass ratios, effective inspiral, and effective precessing spins based on GPU-accelerated binned Gaussian processes~(BGP), we characterize BBH subpopulations with high flexibility, for the first time in four dimensions, and present model-agnostic constraints on their astrophysical origins.
In this \textit{Letter}, using GPU-accelerated binned Gaussian processes~\citep[BGPs,][]{Mohite2022, Mandel:2016prl, Ray:2023upk, Ray:2024hos, Sridhar:2025kvi}, we present the first data-driven reconstruction of the joint four-dimensional distribution of BBH primary masses, mass ratios, effective inspiral, and effective precessing spins. We characterize four subpopulations that span different ranges of primary mass and have distinct distributions of mass ratio, effective inspiral, and effective precessing spin, free of strong model priors~\citep[see also a parallel work by][who approach this problem with strongly modeled parametrizations guided by clusters in the data]{Guttman:2026cnv}. We further find new correlations within specific mass ranges that substantially extend the astrophysical interpretation of these subpopulations, beyond the reach of existing parametrizations. %Key among them is a unique broadening of the effective inspiral spin distribution with effective precessing spin that only appears at high masses~($>45M_{\odot}$) and all mass ratios. By comparing with population synthesis simulations, we show that the data indicate substantial contributions from both hierarchical mergers and massive 1G+1G systems (that are possibly remnants of stellar mergers or BH+star collisions) in this mass range. 

The rest of this \textit{Letter} is organized as follows. In Sec.~\ref{sec:methods} we describe our population model and inference framework. In Sec.~\ref{sec:results} we present our results. In Sec.~\ref{sec:astro} we interpret our results in the context of the astrophysical origins of various BBH subpopulations. In Sec.~\ref{sec:conclusion} we conclude with a discussion of follow-up investigations towards a robust and coherent understanding of BBH formation.
% \section{Key Results}
\section{Population Analysis}
\label{sec:methods}
The observed BBH catalog represents a detectable sample of events drawn from an inhomogeneous Poisson Process~\citep{popgw2,Mandel:2018mve,pop-vitale, popgw3,Thrane:2018qnx}. We model the merger rate density per log primary mass $(m_1)$, mass-ratio $(q)$, redshift $(z)$, and effective spins $(\chi_{eff},\chi_p)$ as a piecewise-binned function of the form:
\begin{equation}
\begin{split}
\frac{dN}{dm_1dqd\chi_{eff}d\chi_pdz}= {} &\frac{n^{\gamma}}{m_1}\frac{dV}{dz}T_{obs}(1+z)^{\kappa-1} \\
       {} & (m_1,q,\chi_{eff},\chi_p)\in\gamma^{th}~\rm{bin}.
\end{split}
\end{equation}
% \begin{equation}
%     \frac{dN}{dm_1dqd\chi_{eff}d\chi_pdz}= \frac{n^{\gamma}}{m_1}\frac{dV}{dz}T_{obs}(1+z)^{\kappa-1}
% \end{equation}
where $V$ is the comoving volume, $T_{obs}$ is the observation time and $\kappa$ the redshift evolution parameter~\citep{Fishbach:2018edt}. The merger rate density in each bin~$(\{n^{\gamma}\}_{\gamma=1}^{\gamma=N_{bins}})$ is a hyperparameter that we constrain from the data using Bayesian hierarchical inference~\citep{Thrane:2018qnx, Mandel:2018mve, pop-vitale, popgw3}. We use a Gaussian process~(GP) prior with an exponential quadratic kernel for the logarithmic rate densities so as to smooth over regions of sparse data~\citep{Foremanmackey2014, Mandel2017, Mohite2022, Ray:2023upk, Ray:2024hos, Sridhar:2025kvi}. The length scales, covariance amplitude, and mean function of the GP are self-consistently inferred along with the rate densities and are themselves modeled using LogNormal, HalfNormal, and Normal priors, respectively~\citep{Mohite2022, Ray:2023upk, Ray:2024hos, Sridhar:2025kvi}.

Note that our population model can, in principle, infer any arbitrary shape and correlation in the four-dimensional BBH population up to the resolution limit imposed by our choice of binning~(see, however, Appendix~\ref{sec:app-sys} for subtleties). While we fix $\kappa$ to the median value~$(\kappa=2.7)$ reported by existing measurements from the same dataset~\citep{LIGOScientific:2026ctl}, we note that varying $\kappa$ is expected to have no significant effect on the inferred shapes~\citep{Ray:2024hos, Sridhar:2025kvi}. We show in Appendix~\ref{sec:app-varry-bink} that our results are robust against reasonable variations in binning choices and $\kappa$ values, and are hence representative of the underlying population independent of model restrictions.

We construct the joint posterior distribution of the rate densities and GP hyperparameters from the LVK's publicly released parameter estimation data of each BBH in GWTC-5 that was detected with a false alarm rate of one per year or lower~\citep{LIGOScientific:2026wfs, LIGO_Scientific_Collaboration_and_Virgo_Collaboration_and_KAGRA_Collaboration2026-is}. We rely on simulated signals injected into detector noise realizations that were ranked and publicly released by the LVK,~\citep{LIGOScientific:2026ctl, LIGO_Scientific_Collaboration2026-vz, Essick:2025Injections} to correct for Malmquist biases arising from the imposition of a stringent detection threshold~\citep{popgw2, Mandel:2018mve, pop-vitale, popgw3}. We monitor the convergence of Monte Carlo sums used to evaluate the hyperposterior~\citep {Pdet1-Farr, Pdet2-essick} by imposing a variance-based penalty on the hierarchical likelihood as proposed by \cite{Talbot2025} and derived by \cite{Sridhar:2025kvi} for binned population models. 

To sample our high-dimensional hyperposterior and mitigate the cubic time complexity of GP-draws~(which was the primary bottleneck of scaling previous implementations beyond three BBH-parameters, \cite{Ray:2023upk, Ray:2024hos, Sridhar:2025kvi}), we use GPU-accelerated Hamiltonian Monte Carlo~\citep[HMC,][]{Brooks_2011} methods. We rely on a \texttt{PyTorch}~\citep{Paszke:2019xhz} based implementation of the No U Turn Sampler~\citep[NUTS,][]{Hoffman2014} available through the \texttt{Pyro} package~\citep{2018arXiv181009538B}. The code developed to implement our analysis is publicly available through the Python package \texttt{gppop-0.1.0}~(\hyperref[https://github.com/AnaryaRay1/gppop]{https://github.com/AnaryaRay1/gppop}). For further details of posterior construction, exact forms of hyper-priors, and scalability with bin resolution, see Appendices~\ref{sec:app-hyperpost} and~\ref{sec:app-scale}, respectively.
\section{GWTC-5: Correlations and Subpopulations}
\label{sec:results}
% \begin{table}
% % \begin{adjustwidth}{0cm}{}
% \begin{tabular}{cc}
% \hline
% \hline
% Parameter & Bin edges \\
% \hline
% $m_1 (M_{\odot})$           & \hspace{1em} log-uniform(5, 200, 25)                                                                              \\

% \hline
% $\chi_{\rm eff}$ & \hspace{-3em} \begin{tabular}[c]{@{}c@{}}-1, -0.85, -0.7, -0.6, -0.4, -0.3, -0.2,   \\ -0.1, -0.05, 0.0, 0.05, 0.1, 0.15,  \\  0.2, 0.3, 0.4, 0.6, 0.7, 0.85, 1
%        \end{tabular}                                                                              \\
% \hline
% $\chi_p$   & \hspace{-2em} \begin{tabular}[c]{@{}c@{}} 0.0, 0.1, 0.15, 0.2, 0.25, 0.3, 0.35,  \\ 0.4, 0.45, 0.5, 0.6, 0.7, 0.8, 0.9, 1.0\end{tabular}  \\
% \hline
% $q$   & \hspace{-2em} uniform(0.1,1,10)\\
% \hline
% \end{tabular}
% % \end{adjustwidth}
% \caption{The bin edges used for the results presented in this section.}\label{tab: Bin choices}
% \end{table} 

\begin{figure*}[t]
    \centering

    \begin{overpic}[
        width=\textwidth
        % Uncomment the following two lines while adjusting the position:
        % ,grid
        % ,tics=10
    ]{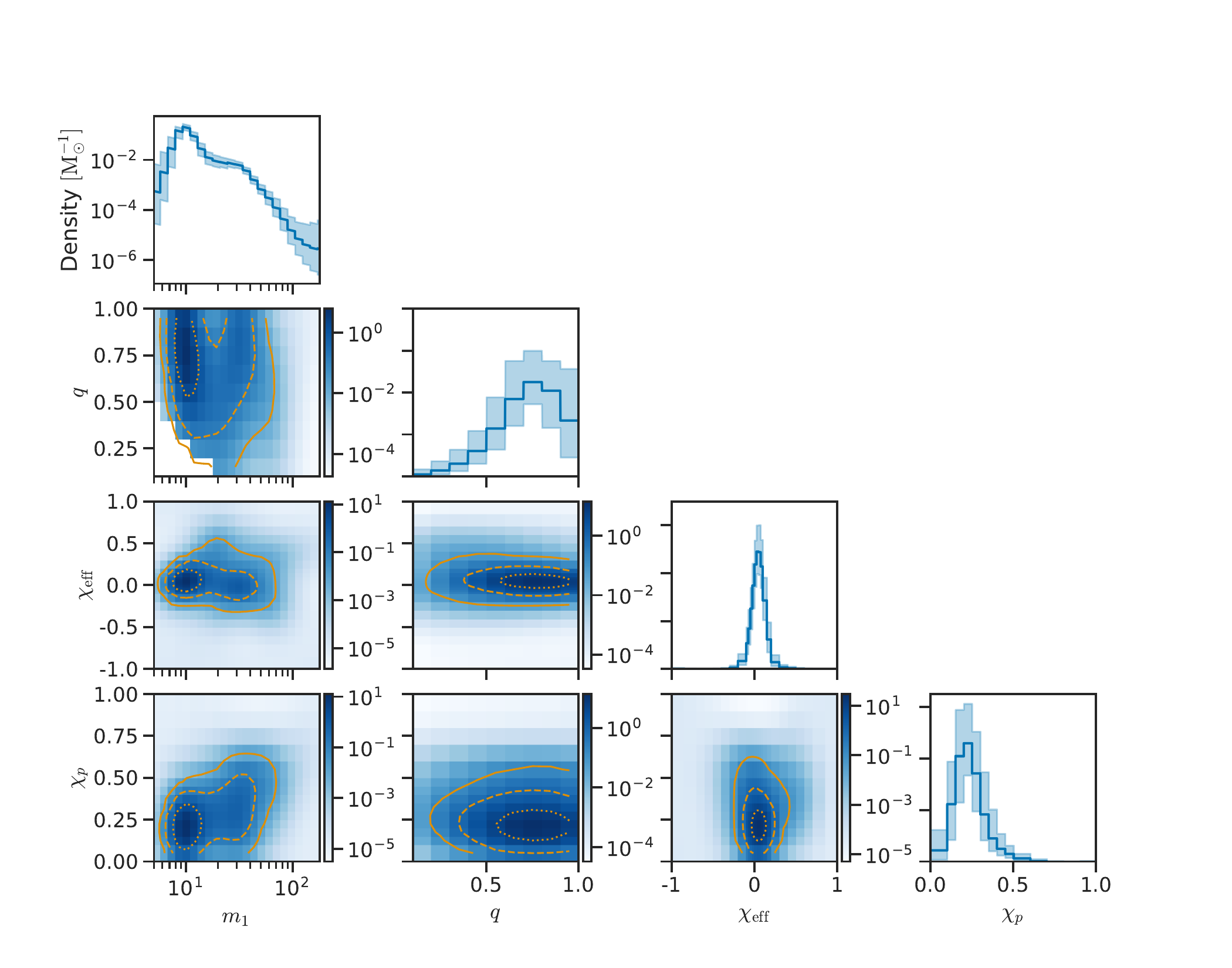}

        % The coordinates (67,58) are percentages of the figure width
        % and height. They specify the lower-left corner of the table.
        \put(54,57){%
            \colorbox{white}{%
                \begin{minipage}{0.45\textwidth}
                    \centering
                    \small
                    \begin{tabular}{cc}
                    \hline
                    \hline
                    Parameter & Bin edges \\
                    \hline
                    $m_1 (M_{\odot})$           & \hspace{1em} log-uniform(5, 200, 25)                                                                              \\
                    
                    \hline
                    $\chi_{\rm eff}$ & \hspace{-3em} \begin{tabular}[c]{@{}c@{}}-1, -0.85, -0.7, -0.6, -0.4, -0.3, -0.2,   \\ -0.1, -0.05, 0.0, 0.05, 0.1, 0.15,  \\  0.2, 0.3, 0.4, 0.6, 0.7, 0.85, 1
                           \end{tabular}                                                                              \\
                    \hline
                    $\chi_p$   & \hspace{-2em} \begin{tabular}[c]{@{}c@{}} 0.0, 0.1, 0.15, 0.2, 0.25, 0.3, 0.35,  \\ 0.4, 0.45, 0.5, 0.6, 0.7, 0.8, 0.9, 1.0\end{tabular}  \\
                    \hline
                    $q$   & \hspace{-2em} uniform(0.1,1,10)\\
                    \hline
                    \end{tabular}
                                     \captionof{table}{The bin edges used for the results presented in this section.}\label{tab: Bin choices}

                \end{minipage}%
            }%
        }

    \end{overpic}

    \caption{Marginal distributions in 1 and 2 dimensions. In the diagonal panels, the shaded regions represent 90\% posterior credible intervals on the marginal 1-dimensional population distributions and the solid lines represent the medians. In the off-diagonal panels, the heat map represents the posterior median on the two-dimensional population density marginalized over the remaining BBH parameters and the orange lines represent the corresponding $95\%,90\%,$ and $50\%$ contours.}
    \label{fig:marginals}
\end{figure*}

From our inferred rate densities, we reconstruct the underlying BBH population given GWTC-5 data upto the chosen bin resolution. We use uniform bins in log primary mass and mass ratio and choose irregular binning for the effective spin parameters to allow for higher resolution in dense regions while retaining computational tractability. The default binning choices are summarized in Table~\ref{tab: Bin choices} and it is shown in Appendix~\ref{sec:app-varry-bink} that alternate binning schemes lead to identical results. In addition to corroborating existing findings, we identify new features and correlations in the population~(marked in bold font), both of which are presented below.

\subsection{Marginal distributions and Broad-Population Correlations}
\label{sec:results-marginal}

In Figure~\ref{fig:marginals}, we show the one- and two-dimensional marginal distributions of BBH parameters. The diagonal panels show the 90\% credible intervals on the posterior of the marginal population and the other panels represent the posterior medians of the two-dimensional densities marginalized over the remaining dimensions. We find that the one-dimensional posteriors are consistent with previous parametric and non-parametric results~\citep{LIGOScientific:2026ctl}. To quantitatively assess the significance of correlations given measurement uncertainties, we further present their fractional uncertainties in Appendix~\ref{sec:app-add}. % For the marginal two-dimensional populations, in addition to displaying median densities, we compute metrics such as the Pearson correlation coefficient~(which quantifies linear correlations between two parameters) and the coskewness (which quantifies change in the conditional variance of one parameter with respect to the other, such as broadening) to quantitatively assess evidence in favour of astrophysically meaningful correlations and present them in Table~\ref{tab:broad-corr-metrics}.

We corroborate the following strongly modeled findings in the marginal distributions and two-dimensional densities:
\begin{enumerate}
    \item{The marginal primary mass density shows two clear features, namely a sharp peak at $10M_{\odot}$ accompanied by a steep decline by $20M_{\odot}$, a flat distribution in the $20-40M_{\odot}$, and steeper fall-off for $m_1>40M_{\odot}$. The marginal mass-ratio distribution is consistent within uncertainties with a flat-distribution in the range~$q\in(0.6,1)$ (even though the median shows a peak near $q\sim0.7$) and a steep fall-off at $q<0.7$. The $\chi_{eff}$ distribution is sharply peaked near small but positive values with no evidence of skewness. The $\chi_p$ distribution peaks bear 0.2 and has a longer positive tail. These results are fully consistent with most previous findings~\citep{LIGOScientific:2026ctl, Sridhar:2025kvi}.}
    \item{In the two-dimensional densities, we find an overall broadening of $\chi_{eff}$ with decreasing $q$ consistent with previous studies~\citep{Vijaykumar:2026zjy, LIGOScientific:2026ctl}, and no evident correlations between $q$ and $\chi_p$. We find four distinct clusters in the $m_1-\chi_{eff}$ plane, some of which correspond to distinct clusters in the $m_1-\chi_p$ and $m_1-q$ planes, corroborating the existence of four distinct mass-based subpopulations reported by past investigations~\citep{Ray:2026uur, Sridhar:2025kvi, Alvarez-Lopez:2026ymo, Flanagan:2026ayy, Flanagan:2026btu, Cheng:2026bpc, Guttman:2026cnv, Banagiri:2025dmy}.}
\end{enumerate}
% In addition, we identify \textbf{new correlations} in the overall population presented below:
% \begin{enumerate}
%     \item{We find a broadening of the $\chi_{eff}$ distribution with increasing $\chi_p$}
%     \item{We find a broadening of the $\chi_{eff}$ with increasing $m_1$ only above $44M_{\odot}$}
%     \item{We find a positive correlation between $\chi_p$ and $m_1$ only above $m_1>44M_{\odot}$}
% \end{enumerate}

Note that any population density marginalized over $m_1$ can in principle be dominated by the subpopulation corresponding to the $10M_{\odot}$ peak, after which, the merger rate density sharply drops by more than an order of magnitude within $12-15M_{\odot}$. The clusters identified in the $m_1-\chi_{eff}$, $m_1-q$, $m_1-\chi_p$ planes motivate in-depth analysis of correlations and population-features in specific slices of mass. Additional information is present in these subpopulations that can further elucidate which formation channels are contributing dominantly in specific mass ranges. Our non-parametric reconstruction of the four-dimensional BBH population facilitates a novel model-agnostic probe into these features whose findings are delineated below.
\subsection{Mass-based transitions in 1 and 2 Dimensions}
\label{sec:results-trans}
\begin{figure*}
    \centering
    \includegraphics[width=0.32\textwidth]{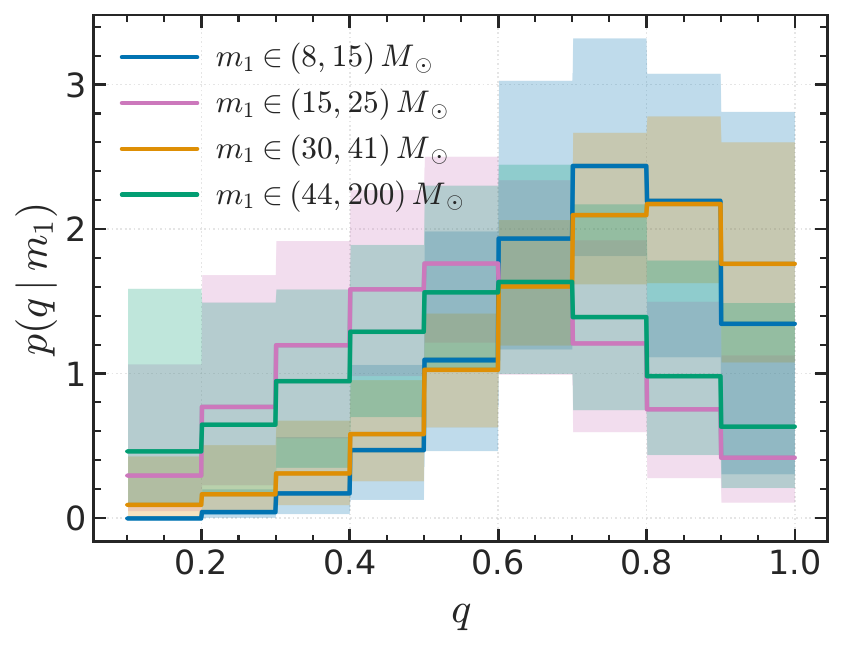}
    \includegraphics[width=0.32\textwidth]{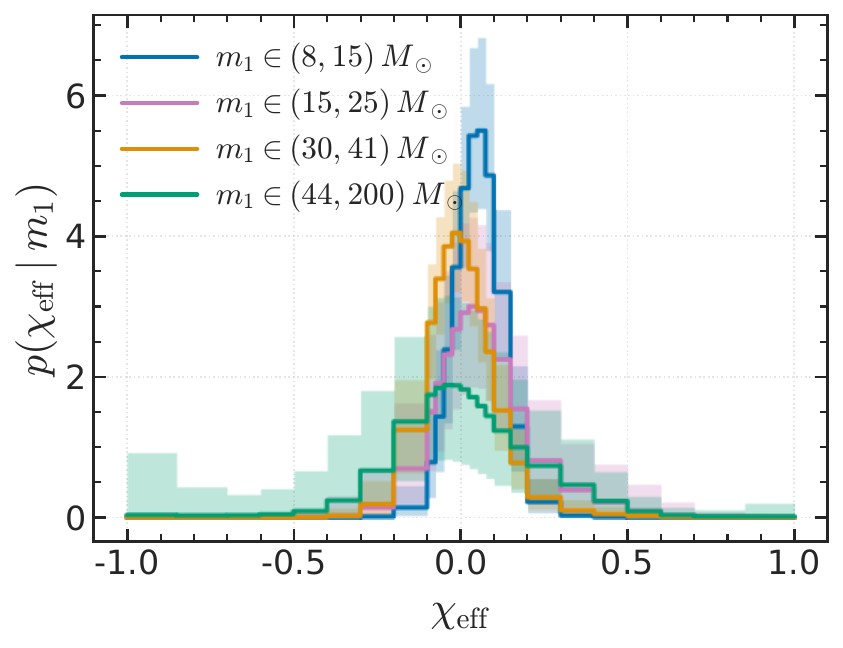}
    \includegraphics[width=0.32\textwidth]{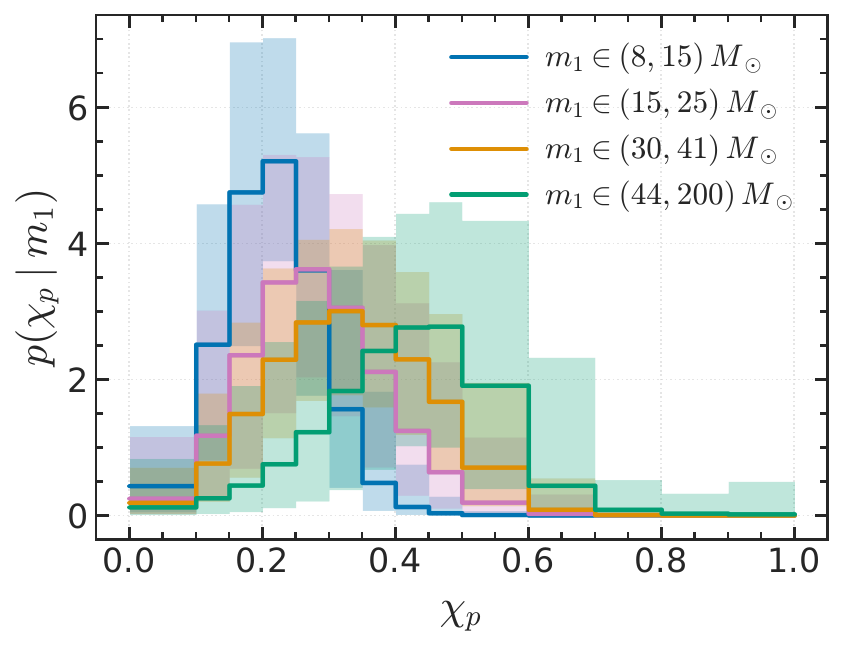}
    \includegraphics[width=0.32\textwidth]{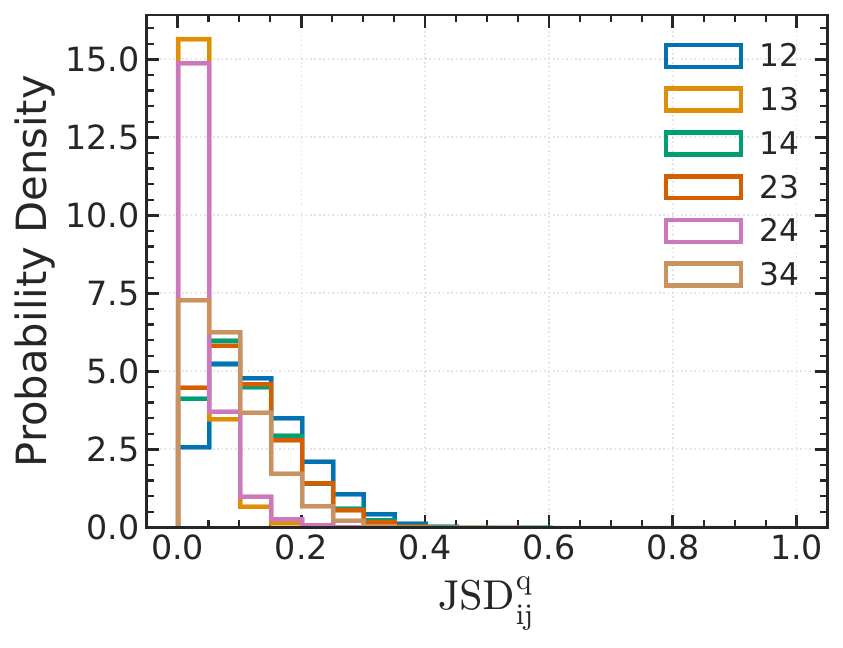}
    \includegraphics[width=0.32\textwidth]{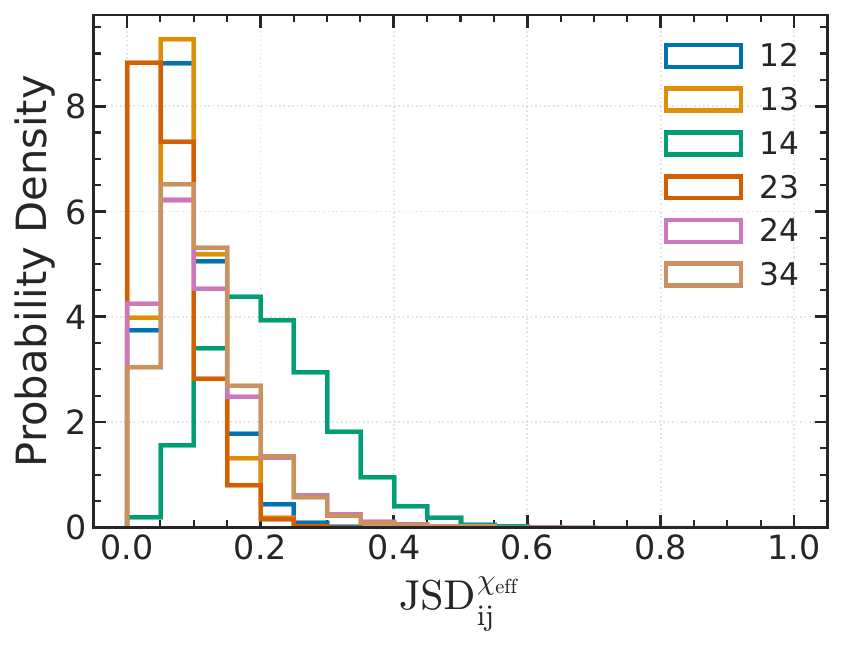}
    \includegraphics[width=0.32\textwidth]{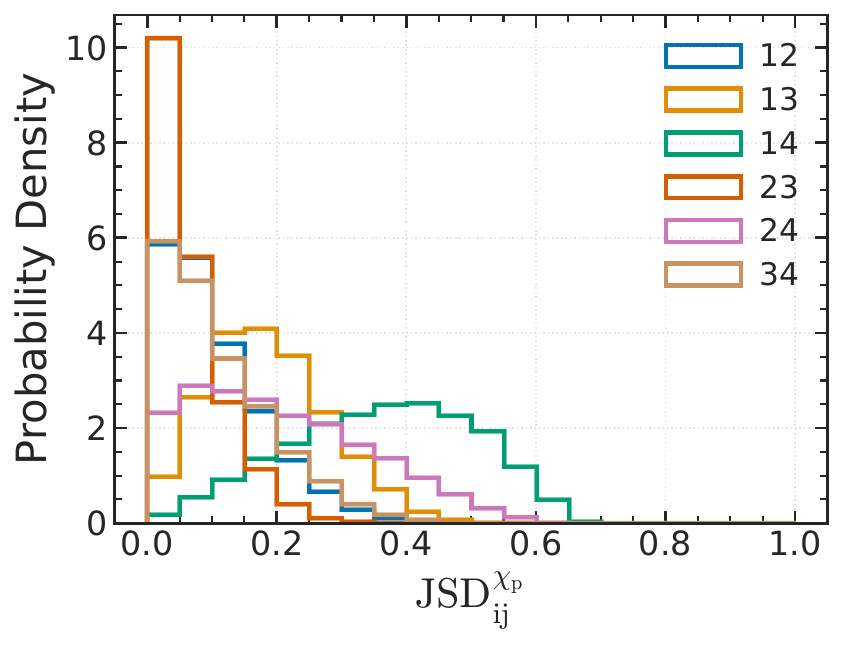}
    \caption{One dimensional distributions conditioned on primary mass and the JS divergences between the distributions corresponding to the different mass-ranges, \label{fig:trans-1D}}
\end{figure*}
\begin{figure*}
    \centering
    \includegraphics[width=0.24\textwidth]{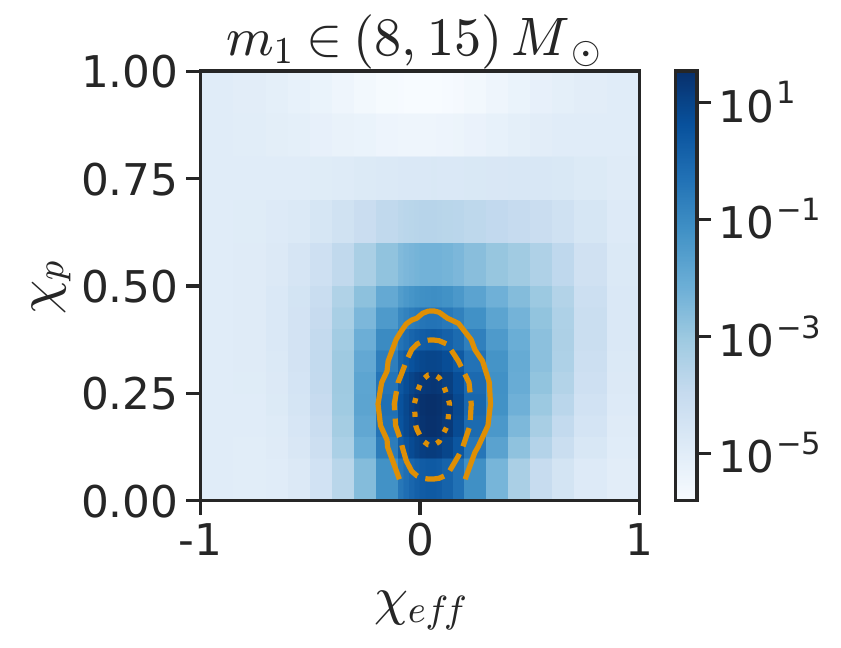}
    \includegraphics[width=0.24\textwidth]{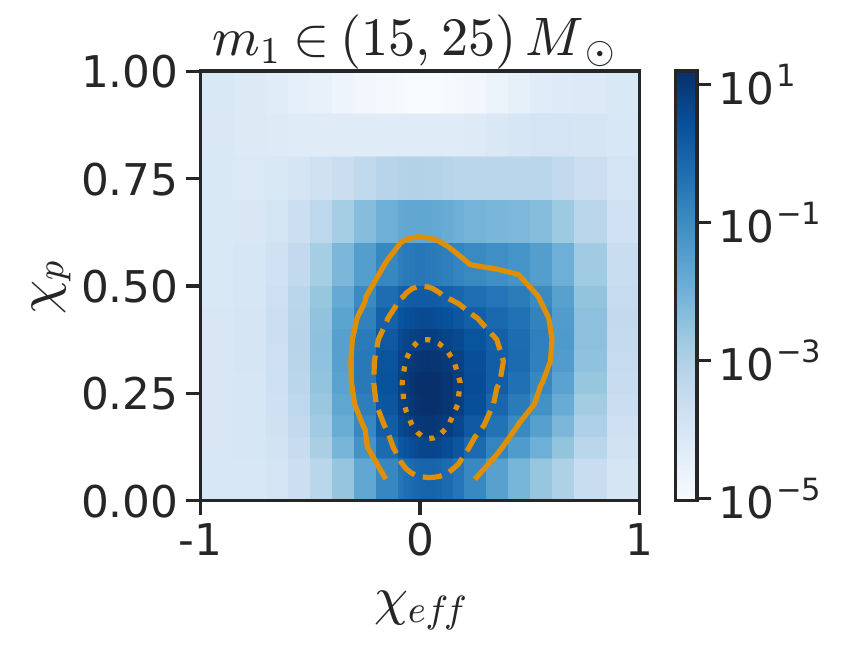}
    \includegraphics[width=0.24\textwidth]{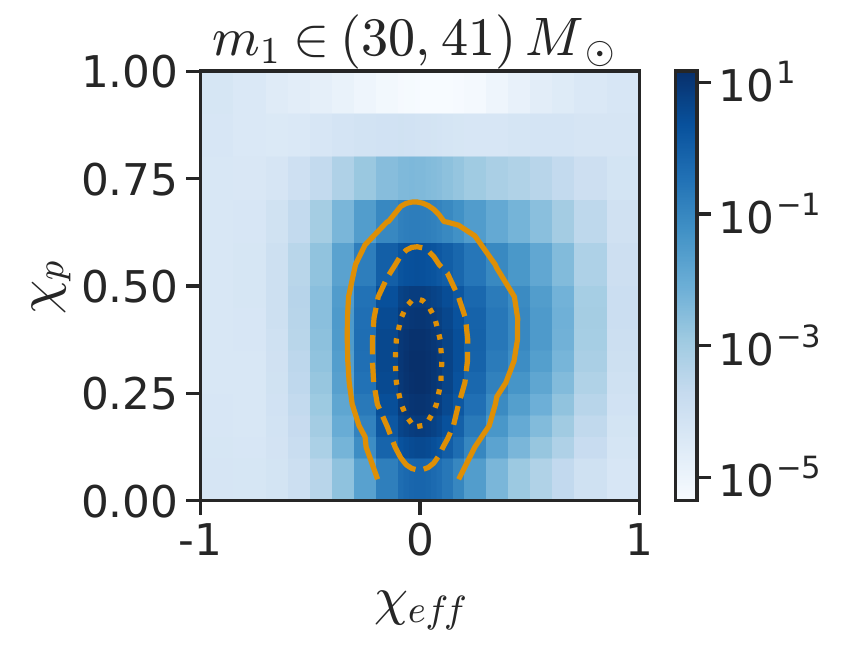}
    \includegraphics[width=0.24\textwidth]{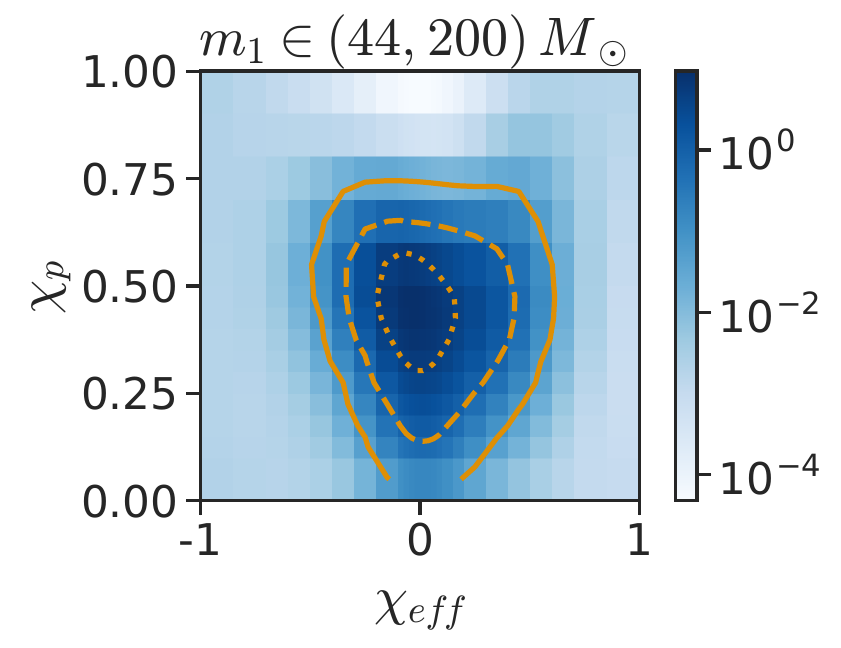}
    \includegraphics[width=0.24\textwidth]{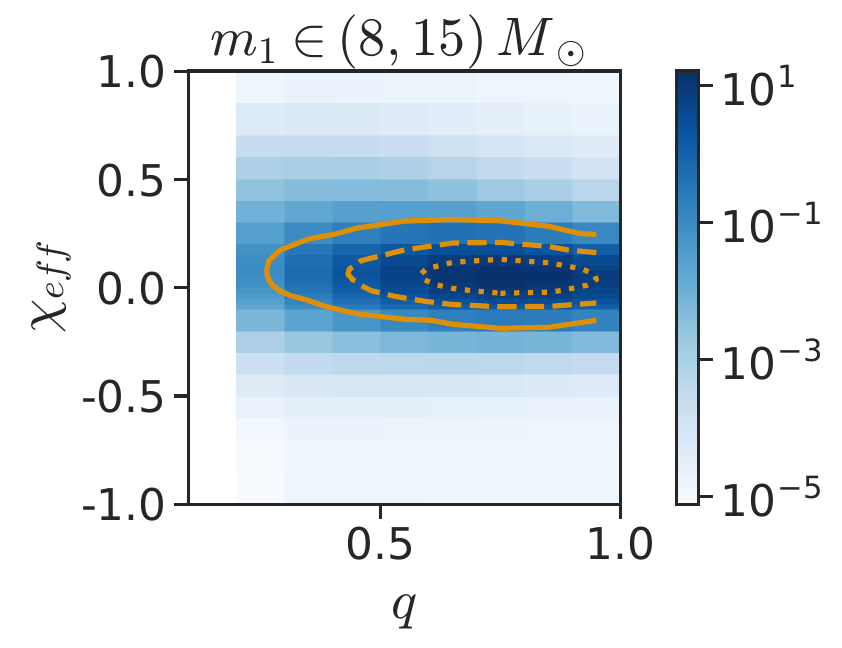}
    \includegraphics[width=0.24\textwidth]{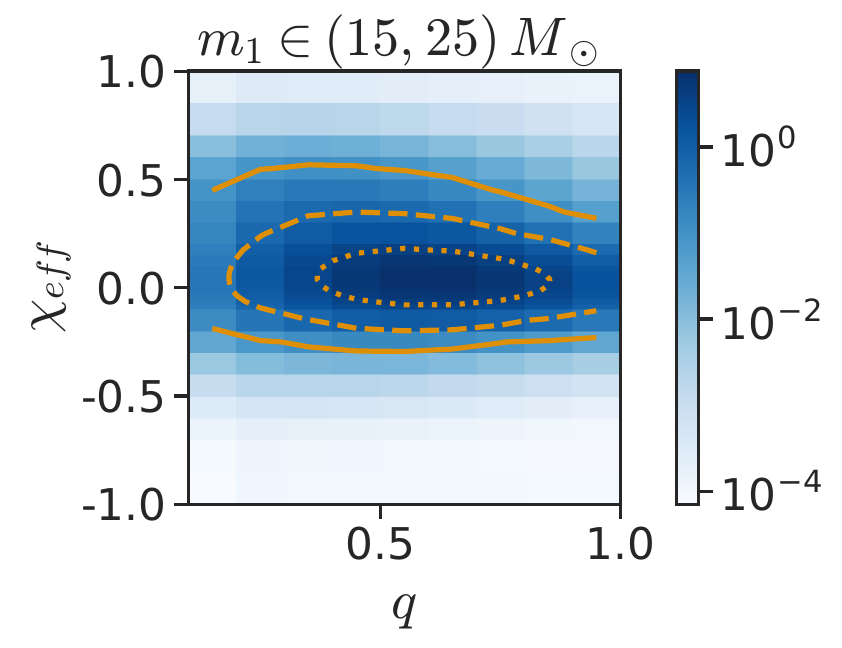}
    \includegraphics[width=0.24\textwidth]{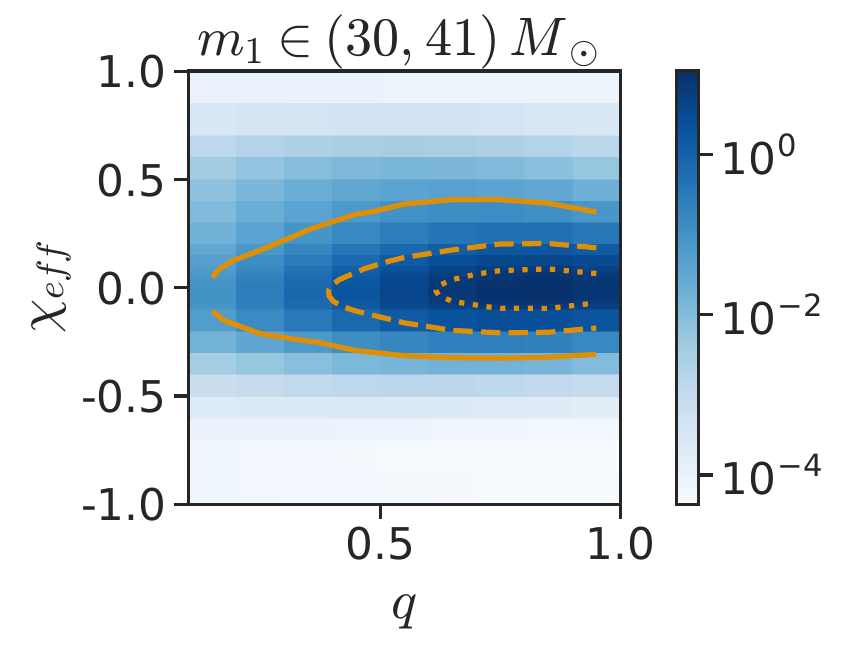}
    \includegraphics[width=0.24\textwidth]{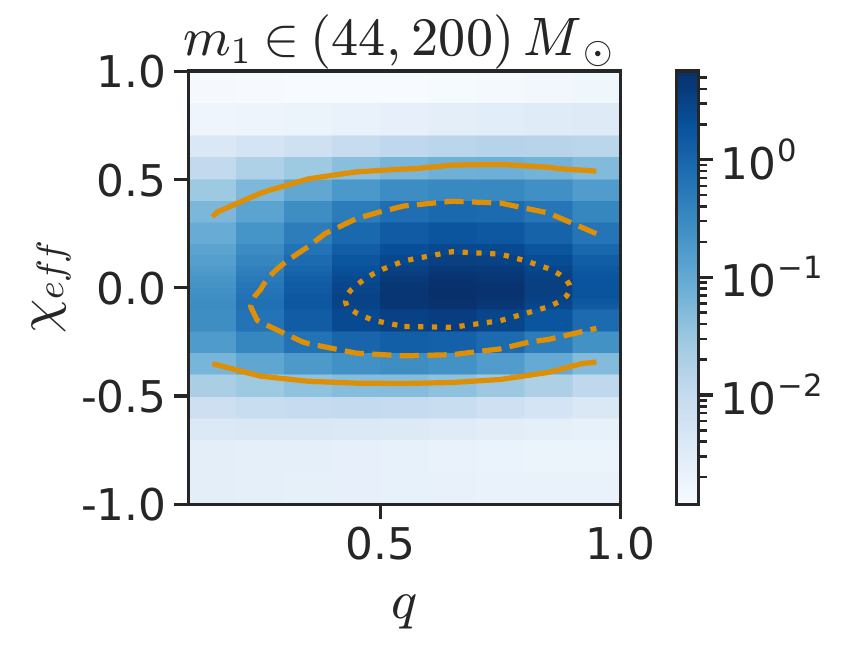}
    \includegraphics[width=0.24\textwidth]{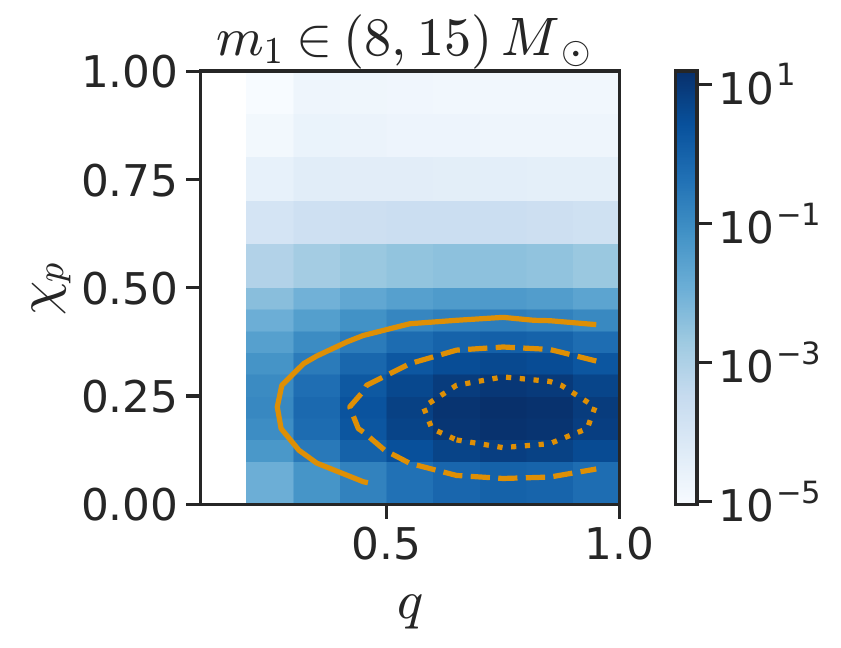}
    \includegraphics[width=0.24\textwidth]{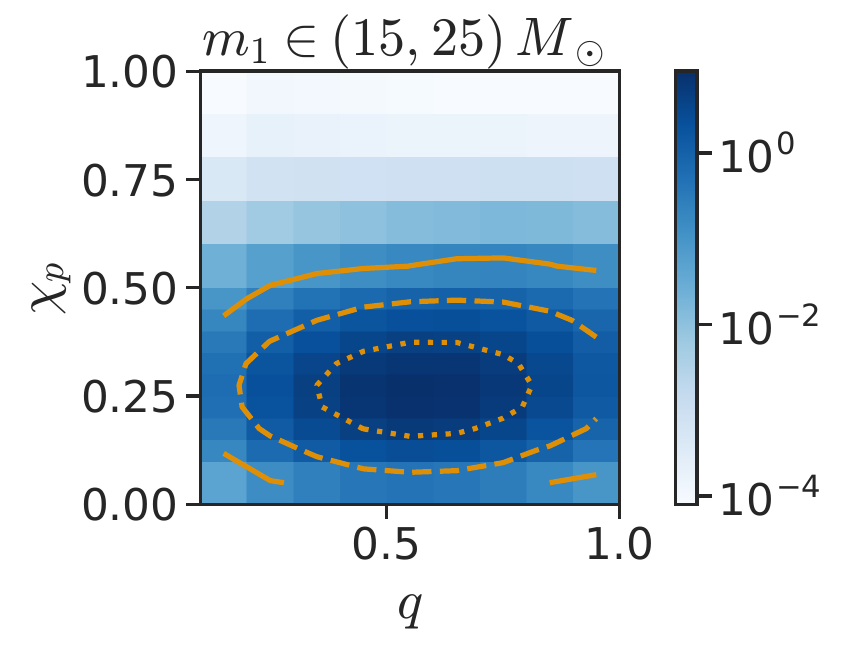}
    \includegraphics[width=0.24\textwidth]{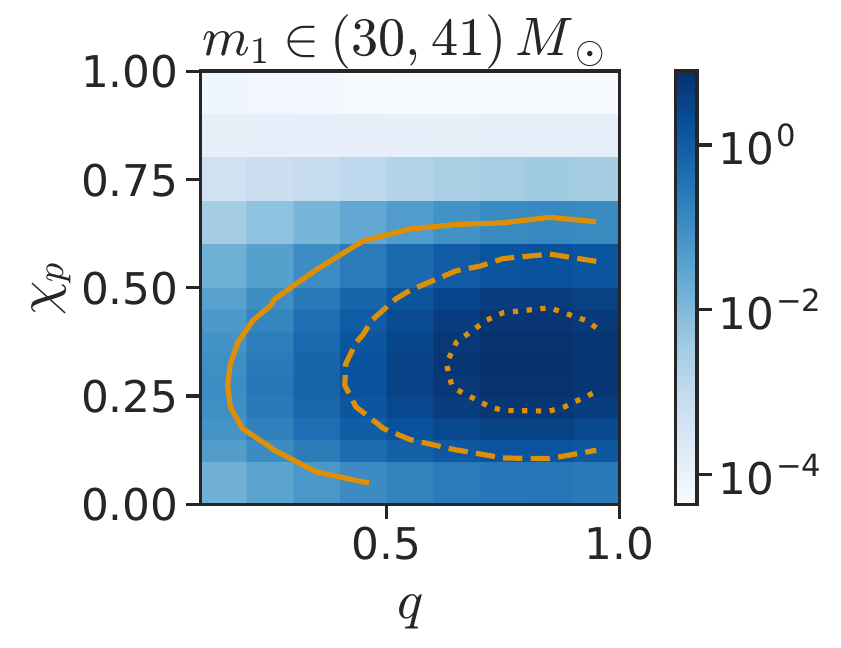}
    \includegraphics[width=0.24\textwidth]{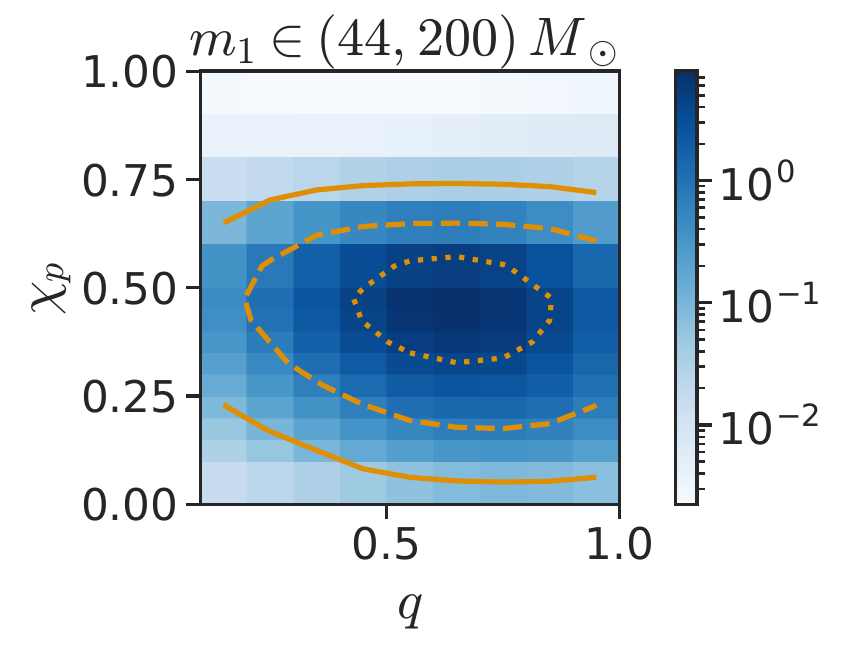}
    \caption{Posterior median of the two-dimensional distributions (density heatmap and the corresponding $95\%,90\%,$ and $50\%$ contours) in different mass-ranges.\label{fig:trans-2D}}
\end{figure*}

To disentangle the relative abundance of different formation channels in specific mass-ranges and to search for additional correlations than the ones driven by the $10M_{\odot}$ peak, we investigate the one- and two-dimensional BBH population in four distinct mass-ranges, that likely correspond to channel-specific subpopulations:
\begin{enumerate}
    \item{Subpopulation 1: $m_1\in (8M_{\odot},15M_{\odot})$}
    \item{Subpopulation 2: $m_1\in (15M_{\odot},25M_{\odot})$}
    \item{Subpopulation 3: $m_1\in (30M_{\odot},41M_{\odot})$}
    \item{Subpopulation 4: $m_1\in (44M_{\odot},200M_{\odot})$}
\end{enumerate}

In addition to reconstructing the one- and two-dimensional populations of BBH parameters in each specific mass-range (shown in Figures~\ref{fig:trans-1D}, and \ref{fig:trans-2D}, respectively), we compute the Jensen-Shannon~(JS) divergence between distributions corresponding to different mass-ranges~(also shown in Figure~\ref{fig:trans-1D}), so as to quantitatively establish the existence of mass-based subpopulations in different slices of the joint distribution of BBH parameters. Furthermore, to quantitatively assess the significance of intrinsic correlations in specific mass-ranges given measurement uncertainties, we present the width of the $90\%$ credible region, in each bin, scaled by their medians in Appendix~\ref{sec:app-add}.%we compute metrics such as the Pearson correlation coefficient~(which quantifies linear correlations between two parameters) and the coskewness (which quantifies change in the conditional variance of one parameter with respect to the other, such as broadening) to quantitatively assess evidence in favour of astrophysically meaningful correlations and present them in Table. We further

\begin{figure*}[t!]
    \centering
    \includegraphics[width=0.24\textwidth]{chieffchip_high.pdf}
    \includegraphics[width=0.24\textwidth]{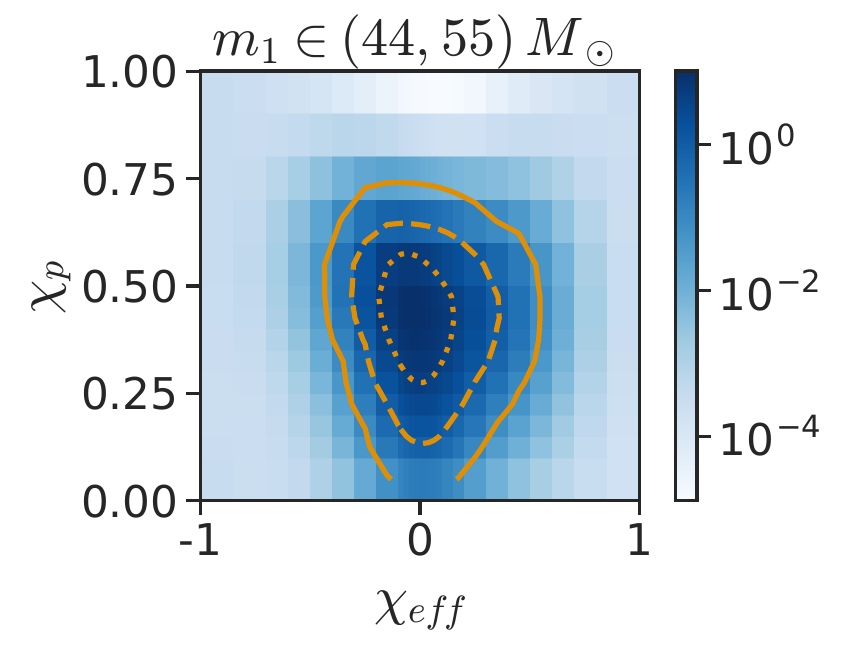}
    \includegraphics[width=0.24\textwidth]{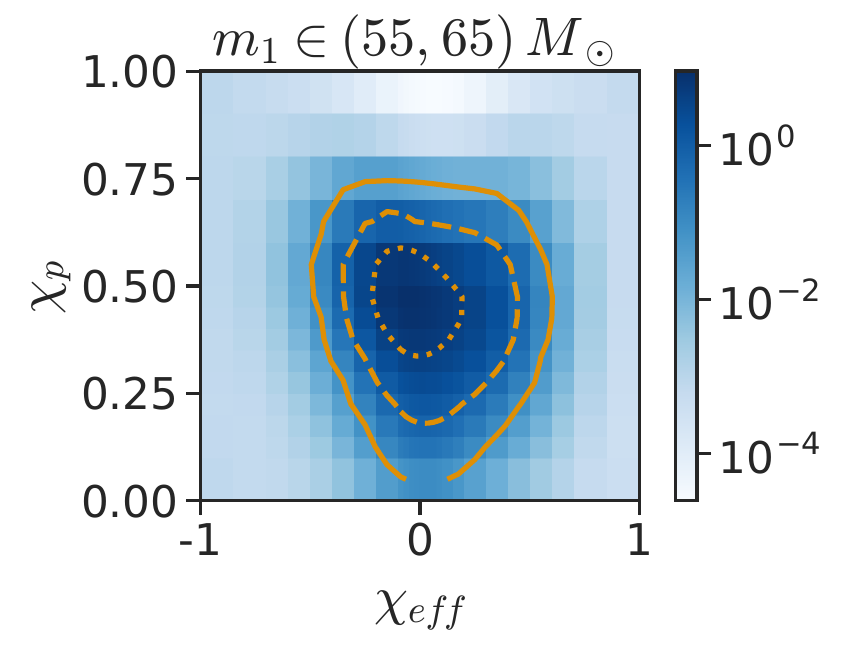}
    \includegraphics[width=0.24\textwidth]{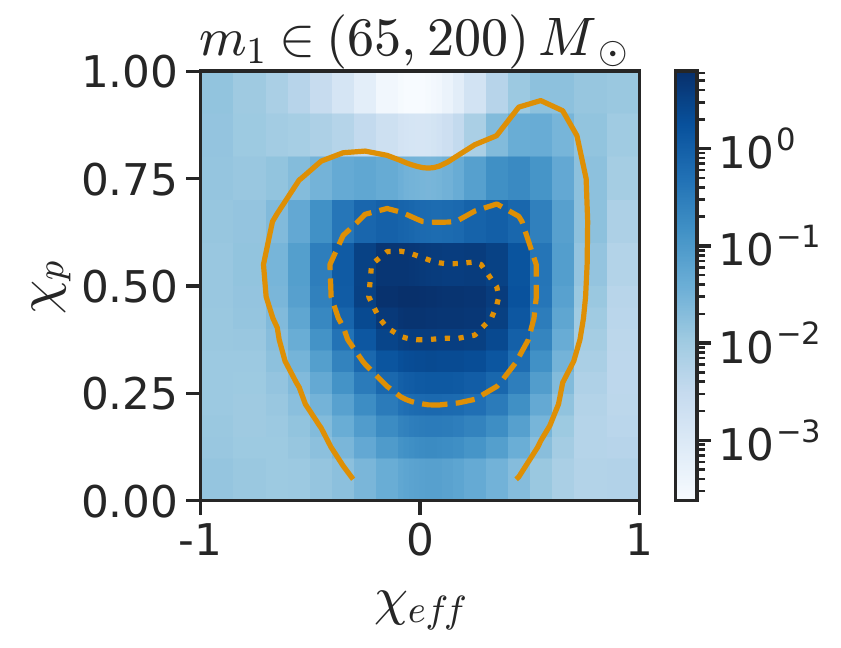}
    \caption{Posterior median of the $\chi_{eff}-\chi_p$ distributions (density heatmap and the corresponding $95\%,90\%,$ and $50\%$ contours) in specific mass ranges above $44M_{\odot}$, \label{fig:trans-2Db}}
\end{figure*}

\begin{figure*}
    \centering
    \includegraphics[width=0.24\textwidth]{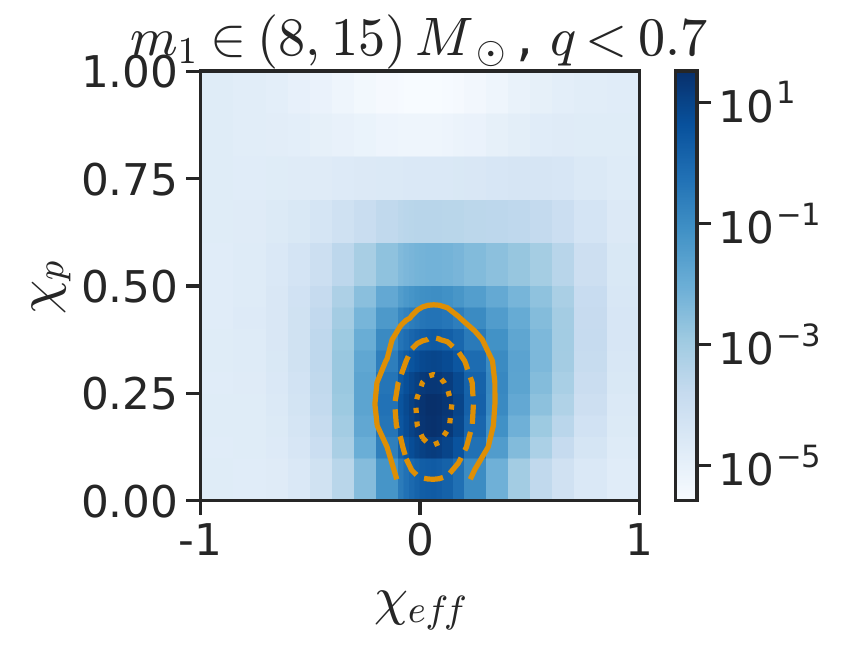}
    \includegraphics[width=0.24\textwidth]{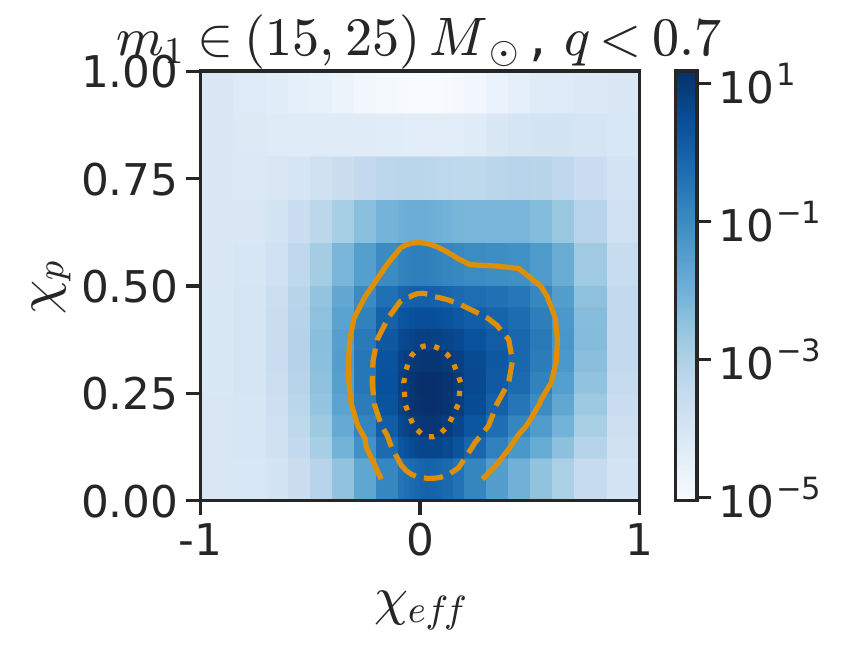}
    \includegraphics[width=0.24\textwidth]{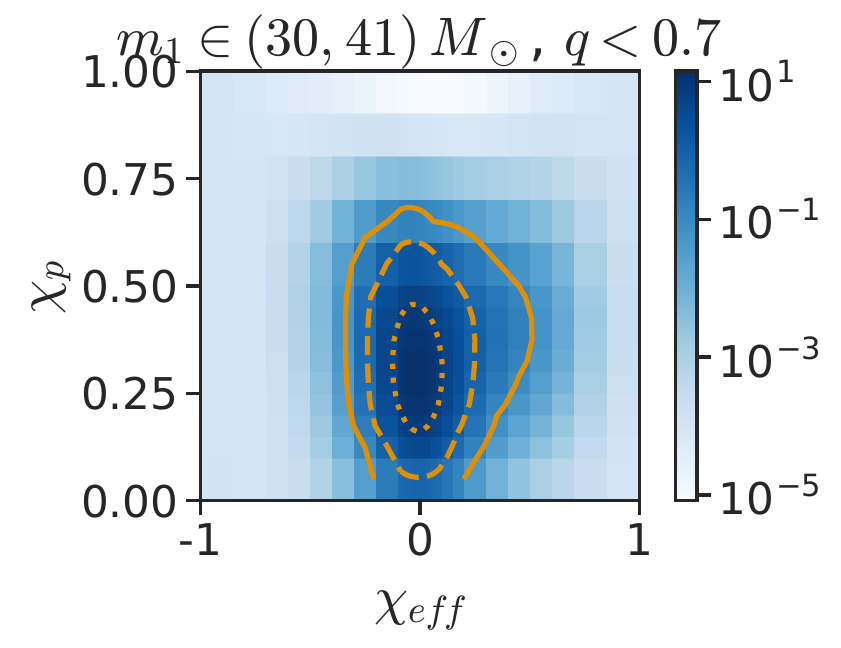}
    \includegraphics[width=0.24\textwidth]{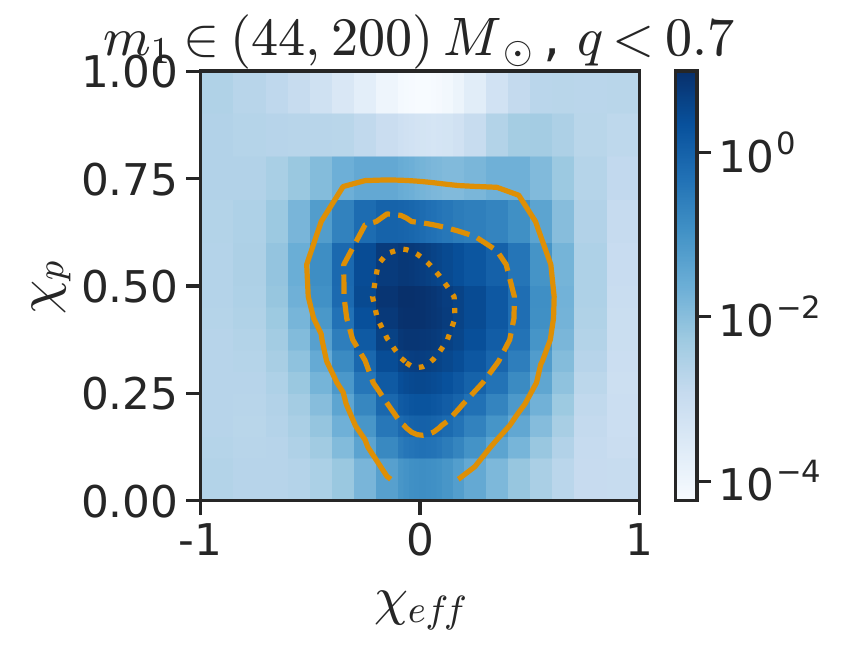}
    \includegraphics[width=0.24\textwidth]{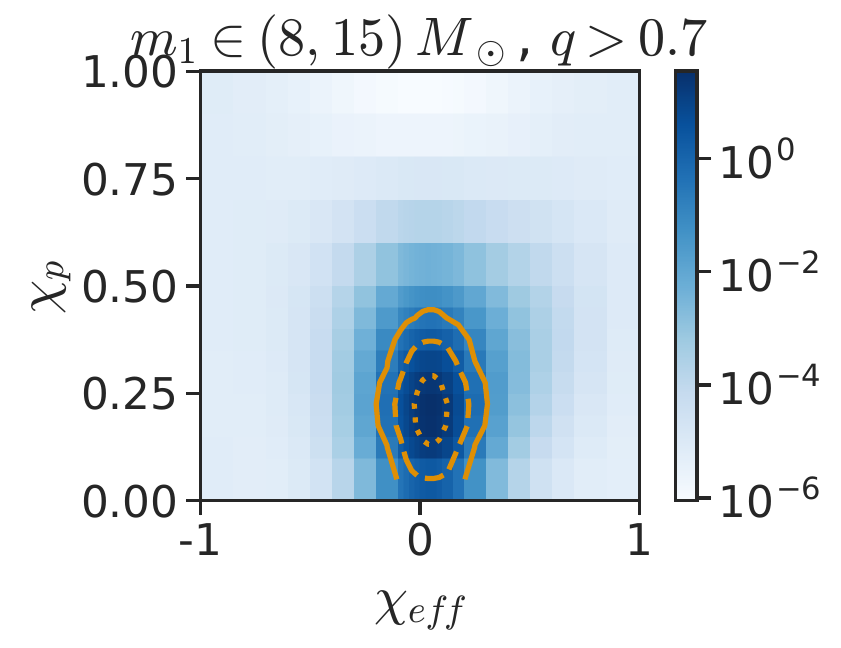}
    \includegraphics[width=0.24\textwidth]{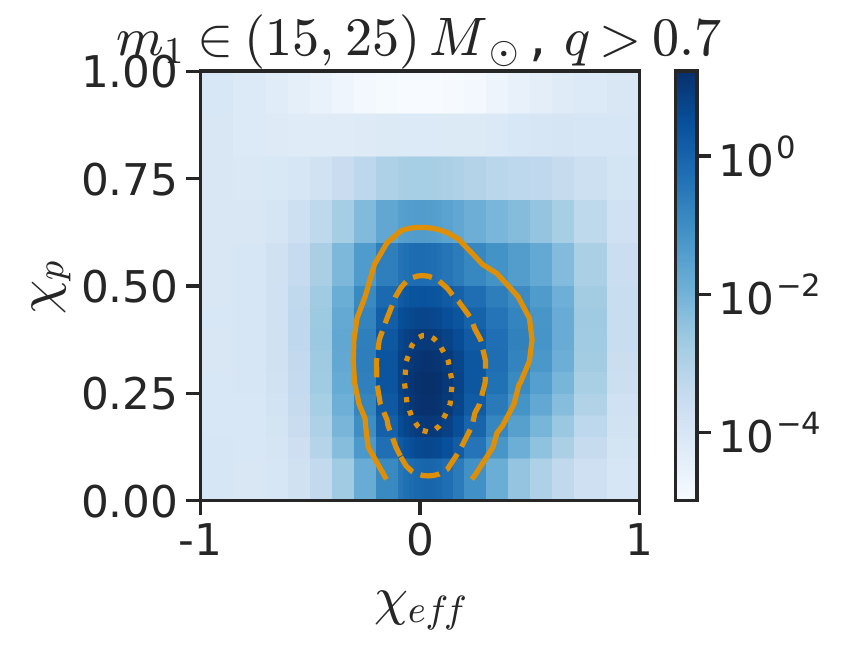}
    \includegraphics[width=0.24\textwidth]{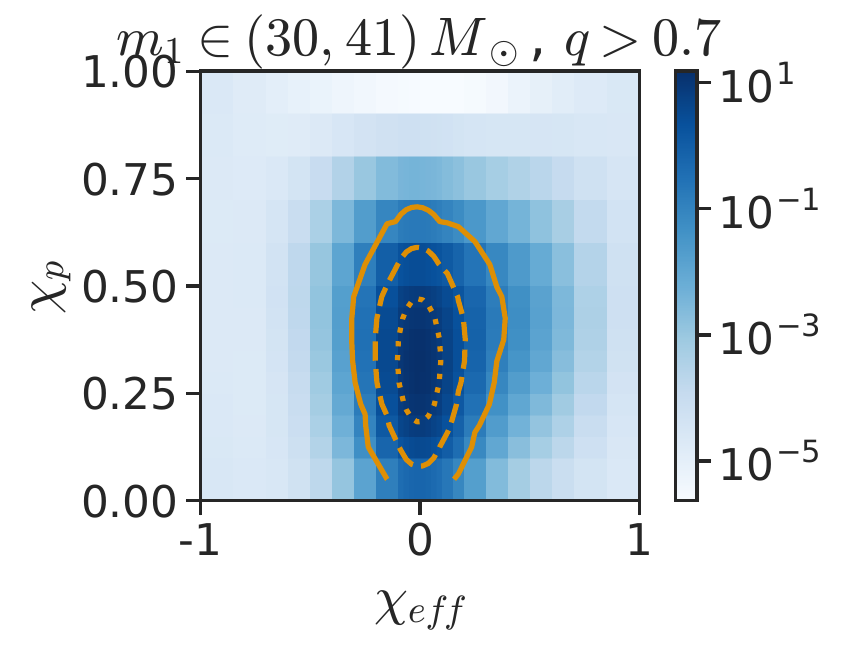}
    \includegraphics[width=0.24\textwidth]{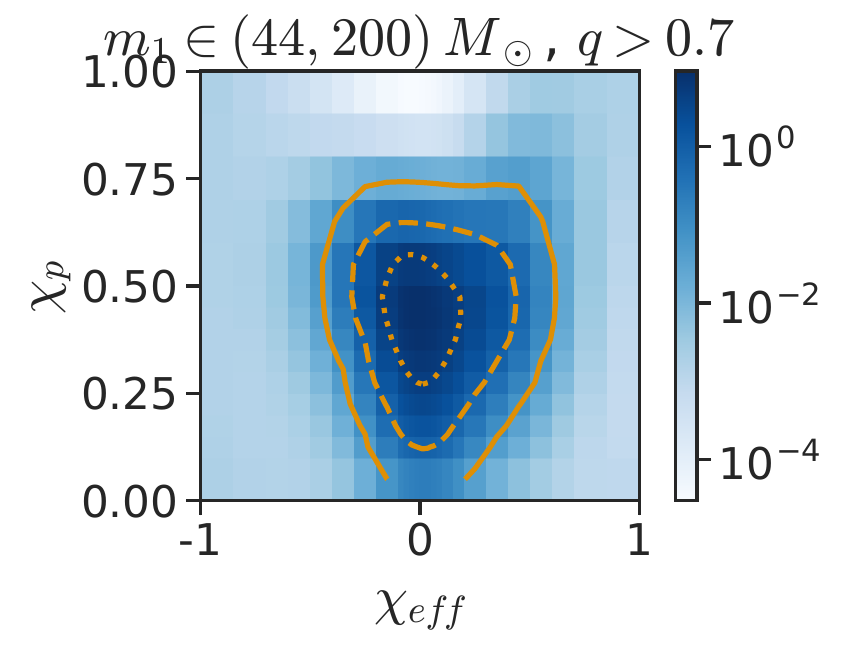}
    
    \caption{Posterior median of the $\chi_{eff}-\chi_p$ distributions (density heatmap and the corresponding $95\%,90\%,$ and $50\%$ contours) in different mass and mass ratio ranges.\label{fig:trans-3D}}
\end{figure*}

From the one-dimensional conditional distributions and JS divergences, we conclude that there is evidence of 4 distinct subpopulations. The two-dimensional distributions in each mass range reveal novel characteristics of these subpopulations. We elaborate the properties of each subpopulation as follows.

\begin{enumerate}
    \item{\textit{Subpopulation 1}, which comprises the $10M_{\odot}$ peak shows preference for small but positive $\chi_{eff}$ values, small $\chi_p$, and a mass-ratio distribution which is nearly flat in the $0.6-1$ range and falls off steeply for $q<0.6$, corroborating the findings of previous strongly-modeled analyses~\citep{Ray:2026uur, Ray:2025xti, Banagiri:2025dmy, Plunkett:2026pxt, Galaudage:2026opk, Cheng:2026bpc}. We find hints of \textbf{a broadening of the $\chi_{eff}$ distribution with decreasing mass ratio in this restricted mass-range.}}
    \item{\textit{Subpopulation 2} which spans the $15-25M_{\odot}$ mass range \textbf{shows a strong preference for unequal mass~$(q\sim 0.5)$ systems}, a positively skewed $\chi_{eff}$ distribution with a tail extending up to high $(\chi_{eff}>0.25)$ values~\citep{Alvarez-Lopez:2026ymo, Rinaldi:2026nyb, Cheng:2026bpc, Flanagan:2026ayy, Flanagan:2026btu}, and \textbf{higher $\chi_p$ values than Subpopulation 1}. We also find a \textbf{broadening of the $\chi_{eff}$ distribution with decreasing mass ratio and a positive correlation between $\chi_{eff}$ and $m_1$ in this restricted mass range.}}%and hints of a positive correlation between $\chi_{eff}$ and $\chi_p$ and an anticorrelation between $\chi_{eff}$ and q
    \item{\textit{Subpopulation 3}, which comprises BBHs in the $30-41M_{\odot}$ range, is consistent with a $\chi_{eff}$ distribution that peaks narrowly and symmetrically at zero, a preference for equal mass ratios and higher $\chi_p$ values than both subpopulations 1 and 2, corroborating previous findings. We find a \textbf{ broadening of the $\chi_p$ distribution with increasing mass ratio and a broadening of the $\chi_{eff}$ distribution with increasing mass ratios, unlike subpopulations 1 and 2.}}
    \item{\textit{Subpopulation 4}, which comprises of high mass~$(m_1>44M_{\odot})$, prefers unequal mass ratios but has higher support at $q>0.7$ than subpopulation 2, has a broad $\chi_{eff}$ distribution symmetric about zero and prefers higher $\chi_p$ values than all other subpopulations. \textbf{There is evidence of a broadening of the $\chi_{eff}$ distribution with increasing $\chi_p$ in this mass range alone and no significant correlation between the spin parameters and mass ratio.} We do not find a preference for positive $\chi_{eff}$ values at high masses in contrast to strongly modeled investigations \citep{Li:2025iux, Rinaldi:2026nyb, Cheng:2026bpc}. Furthermore, this subpopulation shows \textbf{broadening of $\chi_{eff}$ distribution with increasing $m_1$ and a positive correlation between $\chi_p$ and $m_1$ }}
\end{enumerate}

We note here that the unique $\chi_{eff}$ broadening with $\chi_p$ in the high mass subpopulation is a novel property revealed by our model-agnostic analysis which has interesting astrophysical implications. However, it is possible that this correlation might be resulting from marginalizing over intrinsic correlations between $\chi_{eff}-q$ and $\chi_{p}-q$ or $\chi_{eff}-m_1$ and $\chi_{p}-m_1$, which we saw in Figure~\ref{fig:marginals}. Further investigation is necessary to establish whether or not this is the case.

In Figure~\ref{fig:trans-2Db}, we compare the posterior median of the $\chi_{eff}-\chi_p$ population density at specific mass slices above $44M_{\odot}$, namely, $m_1\in(44, 55)$, $m_1\in(55, 65)$ and $m_1\in(65, 200)$, respectively with the corresponding distribution in the entire high mass range of $m_1\in(44, 200)$. We find that \textbf{the broadening of $\chi_{eff}$ with increasing $\chi_p$ is present in every specific mass slice and therefore conclude that this novel characteristic of the high-mass subpopulation is not emerging from marginalization over intrinsic $\chi_{eff}-m_1$ and $\chi_p-m_1$ correlations that are also present in this mass-range}. However, there is still a possibility that this broadening results from the contributions of mass ratio specific subpopulations each with distinct $\chi_{eff}$ and $\chi_p$ distributions.

\subsection{Higher dimensional properties}
\label{sec:results-4D}

Key among the results presented so far is the unique mass-dependent $\chi_{eff}-\chi_p$ correlation. We now explore whether or not this correlation emerges from a mixture of mass-ratio specific subpopulations. In Figure~\ref{fig:trans-3D}, we present the two-dimenstional $\chi_{eff}-\chi_p$ distribution conditioned on different ranges of $m_1$ and $q$. In particular, for each mass-range defined in the previous section, we compute the $\chi_{eff}-\chi_p$ distribution for low $(q<0.7)$ and high $(q>0.7)$ mass-ratio values, and find that both mass ratio ranges show identical features in the $\chi_{eff}-\chi_p$ plane. Hence, we conclude that the $\chi_{eff}-\chi_p$ broadening observed in the high-mass population is not a result of marginalizing over intrinsic $q-\chi_{eff}$ and $q-\chi_p$ correlations and is in fact in intrinsic property of the underlying population.%We further show the $90\%$ posterior credible intervals on the conditional distribution $p(\chi_{eff}|\chi_p, q, m_1)$ for various ranges of the conditioned variables and also compute the JS divergence between the distributions corresponding to the different ranges in Figure. 

\section{Astrophysical Interpretation}
\label{sec:astro}
%Together, this generic characterization of BBH subpopulations presents a novel astrophysical scenario of BBH formation given GWTC-5, that substantially extends the conclusions of previous strongly modeled investigations. 
% Our model-agnostic reconstruction of the joint population distribution of BBH primary mass, mass ratio, effective inspiral and effective precessing spin parameters reveals new correlations and subpopulations and presents a data-driven corroboration of previously reported features found through strongly modeled approaches. 
The interpretation of our results strongly relies on features in the inferred effective spin and mass-ratio distributions, which map directly to astrophysically implicative component spin and component mass properties. In previous works~\citep{Ray:2026uur}, effective spin and mass-ratio features in specific mass-ranges were used to identify which formation channels contribute dominantly in the corresponding subpopulations. Our inferred features and correlations, both in the overall population as well as in particular slices of primary mass, reveal additional details regarding the imprints of specific sub-channels in each mass-range, implicating a substantial extension of the three-component interpretation of \cite{Ray:2026uur}. We summarize this below.

\subsection{The $10M_{\odot}$ peak}
%This subpopulation, owing to its preference for systems with aligned slowly spinning components, remains a strong candidate for isolated binary evolution~\citep{Kalogera:1999tq, Bavera:2020inc, Gerosa:2018wbw, Steinle:2022rhj} even though substantial contamination from other formation scenarios might be present. In particular, the non-negligible fraction of anti-aligned~$(22.4^{+11.5}_{-9.7}\%)$ and in-plane systems (support for $\chi_p>0$) might indicate a comparable presence of 1G+1G dynamical systems. A conservative estimate of the fraction of dynamical systems in this subpopulation would be $(44.7^{+23.0}_{-19.5}\%)$, under the assumption that isolated binary evolution leads to negligible contribution in the $\chi_{eff}<0$ population~\citep[see, however,][for the possibility of large natal kicks imparted by the supernovae during BH formation, which can in-turn increase the fraction of anti-aligned systems in isolated binary evolution]{Wysocki:2017isg, Callister:2020vyz, Fragione:2021qtg}. 
This subpopulation, owing to its preference for systems with aligned slowly spinning components~(narrow $\chi_{eff}$ distribution peaking at small positive values, Figure~\ref{fig:trans-1D}), is consistent with isolated binary evolution~\citep{Kalogera:1999tq, Bavera:2020inc, Gerosa:2018wbw, Steinle:2022rhj}. However, the non-negligible fraction~$(22.4^{+11.5}_{-9.7}\%)$ of anti-aligned~($\chi_{eff}<0$) and in-plane systems ($\chi_p$ peaks at $\sim0.2$, Figure~\ref{fig:trans-1D}) might either indicate large natal kicks in an exclusively isolated scenario~\citep{Kalogera:1999tq} or a comparable presence of 1G+1G dynamical assembly in dense clusters~\citep{Mapelli:2021gyv, Chattopadhyay:2023pil, Rodriguez_2022} or triple systems~\citep{Antonini:2017ash, Rodriguez:2018jqu, Liu:2018nrf, Stegmann:2025zkb}, corroborating previous findings~\citep{Ray:2026uur}.

However, the broadening of $\chi_{eff}$ with decreasing $q$ in this specific mass-range hinted in our results~(Figure~\ref{fig:trans-2D}), if corroboated with higher signifcance from future (larger) catalogs and strongly modeled parametrizations, would implicate,  \textbf{a small but non-negligible contribution from hierarchical mergers \citep[either from clusters or from AGN disks in this subpopulation,][]{Vijaykumar:2026zjy, Gerosa:2021mno} in this specific mass-range}. This in turn implies the existence of 1G+1G mergers near $\sim5M_{\odot}$,  potentially at higher redshifts, which is likely beyond our current detector horizon at that mass-range.

Even though we find a sharp drop-off in the merger-rate density (by more than an order of magnitude within $12-15M_{\odot}$, Figure~\ref{fig:marginals}, \cite{Farah:2023Bump, Tiwari:2021Features, Godfrey:2023Cosmic, Adamcewicz:2024Dip}), we do not find any signature of a gap in this mass range, in-contrast to some strongly modeled investigations~\citep{Legred:2026oiz, Galaudage:2024meo, Galaudage:2026opk, Tiwari:2020otp, Tiwari:2025lit}. It is possible that a gap narrower than our bin resolution exists in the underlying population or that the strongly-modeled analyses are prior-driven. If this feature exists along with the peak, it can be interpreted as evidence for increased compactness of low-mass progenitor cores~\citep{Schneider:2020vvh, Galaudage:2024meo, Galaudage:2026opk} and the possibility of failed supernovae leading to efficient formation of $10M_{\odot}$ BHs through direct collapse~\citep{Legred:2026oiz}.

\subsection{The $15M_{\odot}-25M_{\odot}$ range}
The preference for aligned and rapidly spinning BH components (large positive $\chi_{eff}$ values, Figure) and highly unequal mass~($q$ peaks at $\sim 0.5$), systems in this mass-range~(Figure \ref{fig:trans-1D}) can be indicative of hierarchical mergers in AGN disks[and not dense star clusters which would prefer isotropy,][]~\citep{Tagawa:2020dxe, McKernan:2021nwk, Li:2022cul,Mckernan:2017ssq, Santini:2023ukl, McKernan:2023xio, McKernan:2024kpr, Cook:2024ajp, Fabj:2025vza, Rodriguez:2019huv} dominating this subpopulation, even though subchannels of isolated binary evolution that undergo super-Eddington accretion during mass-transfer onto the first-born BH~\citep{Briel:2022cfl, vanSon:2020zbk} or comprise tidally spun-up components~\citep{Olejak:2021iux, Olejak:2024qxr} can also explain these properties. Previous strongly modeled analyses~\citep{Alvarez-Lopez:2026ymo, Flanagan:2026ayy, Flanagan:2026btu, Cheng:2026bpc} have been unable to distinguish whether only one or both of these channels are present.

Our results, specifically the broadening of $\chi_{eff}$ with decreasing mass-ratio in this specific mass-range~(Figure~\ref{fig:trans-2D}) implicate that \textbf{hierarchical mergers in AGN disks cannot be the only channel contributing to this subpopulation, since broadening requires a mixture of hierarchical channels along with one that produces more equal-mass slowly spinning systems}~\citep{Vijaykumar:2026zjy}. The positive correlation between $\chi_{eff}$ and $m_1$, and no correlation between $m_1$ and $q$ in this specific mass range~(Figures~\ref{fig:marginals}) also indicates that exclusive contributions from the mentioned sub-channels of isolated binary evolution are unlikely, since higher accretion rates would not only increase both $m_1$ and $\chi_{eff}$ but also decrease $q$. Comparable contributions from hierarchical mergers in AGN disks and sub-channels of isolated binary evolution are fully consistent with our results.  %Furthermore, the positive correlation between $m_1$ and $\chi_{eff}$ is not traditionally expected in an exclusively hierarchical subpopulation but can be explicable either through the in this specific mass-range can be an indication  %On the other hand, the preference for high $\chi_p$ in this mass range might also indicate substantial contribution from triples~\citep{Antonini:2017ash, Rodriguez:2018jqu, Liu:2018nrf, Stegmann:2025zkb} even though triples would be only able to explain either subpopulation 1 or 2 but not both.

\subsection{The $30-40M_{\odot}$ range}
This subpopulation prefers slowly spinning components with equal contributions from aligned and anti-aligned orientations~(narrow $\chi_{eff}$ distribution symmetric about zero), equal mass systems~$(q\sim 1)$, and a higher fraction of in-plane orientations~(higher $\chi_p$ mode) than subpopulations 1 and 2~(Figure~\ref{fig:trans-1D}), which is highly consistent with 1G+1G dynamical assembly in dense star clusters~\citep{Ray:2024hos, Sridhar:2025kvi, Ray:2026uur}. The preference for equal masses, small spin magnitudes, and isotropy implies negligible contributions from hierarchical mergers and isolated binary evolution ~\citep{Roy:2025MidThirties}, respectively.  The broadening in the $\chi_p$ and $\chi_{eff}$ distributions with increasing mass-ratio in this specific mass-range~(Figure~\ref{fig:trans-2D}) can be indicative \textbf{of additional formation channels contributing substantially to this subpopulation} such as triple systems~(due to the overall preference for high $\chi_p$). Note, however, that an abundance of 1G+1G systems in this mass range might also indicate a peak in the mass-spectrum near $60-80M_{\odot}$ due to higher generation mergers~\citep{Ginat:2026awh}, which we do not recover with any statistical significance from the current dataset.
\subsection{The High Mass $(>44M_{\odot})$Subpopulation}

This subpopulation shows the highest contribution from systems with large spin magnitudes~(broad $\chi_{eff}$), a preference for equal abundance of aligned and anti-aligned systems~($\chi_{eff}$ symmetric), asymmetric mass ratios, and the highest fraction of in-plane systems~(high $\chi_p$ mode, Figure~\ref{fig:trans-1D}). These features are consistent with substantial contribution from hierarchical mergers~\citep{Fishbach:2017BigBlackHoles} in dense star clusters (and not AGN disks, due to a preference for isotropy) in this subpopulation and suppressed contribution from isolated binaries and triple systems. However, the significant fraction~$(30.6^{+21.2}_{-15.4}\%)$ of systems in the $q>0.7$ range and the broadening of $\chi_{eff}$ with $\chi_p$~(Figure~\ref{fig:trans-2D}) \textbf{is inconsistent with the hypothesis that this subpopulation is exclusively hierarchical in origin, which would require a robust eliptical feature in the $\chi_{eff}-\chi_p$ plane~\citep{Plunkett:2026pxt, Payne:2024ywe}, and is indicative of substantial abundance of other dynamical processes~(due to a preference for isotropy)}. Furthermore, the broadening of $\chi_{eff}$ with increasing $m_1$ and the positive $\chi_p-m_1$ correlation in this specific mass-range~(Figure~\ref{fig:marginals}), which persists up to $\sim70-80M_{\odot}$ \textbf{indicates that these other channels continue to contribute up to such high masses albeit with decreasing abundance relative to hierarchical mergers as a function of $m_1$.}

% \begin{figure}
%     \centering
%     \includegraphics[width=0.9\linewidth]{pq_given_m1_highcmc.pdf}
%     \includegraphics[width=\linewidth]{chieffchip_high_cmc50.pdf}
%     \includegraphics[width=\linewidth]{chieffchip_high_cmc100.pdf}
%     \caption{Comparing our inferred $q$ (top) and $\chi_{eff}-\chi_p$~(center and bottom) distribution for $m_1>44M_{\odot}$ with theoretical simulations of BBH formation in globular clusters. The models with 1G+1G mergers that comprise remnants of BH+star collisions\textsuperscript{\ref{fn:fulya}} (with accretion efficiency $50\%$ on top and $100\%$ at the bottom panel) were taken from \cite{Kiroglu:2024xpc}. }
%     \label{fig:chieffchiphighcmc}
% \end{figure}
The inferred $\chi_{eff}-\chi_p$ distribution in this mass-range likely indicates \textbf{substantial contributions from two isotropic subpopulations, with high and moderate spin magnitudes, respectively}. In particular, 1G+1G dynamical assembly in dense star clusters that comprises components that are remnants of BH+star collisions~\citep{Kiroglu:2024xpc} and/or stellar mergers~\citep{Satish:2026ldf} is likely to be present in this mass range in comparable abundance to hierarchical mergers, giving rise to the observed features. Further investigations and comparison with theoretical models are necessary to verify this interpretation. Nevertheless, the lack of evidence in favour of an exclusively hierarchical contribution in this mass-range indicates that current data show no concrete feature consistent with the onset of the pair instability cut-off at $\sim 45M_{\odot}$~\citep{Ray:2025xti, Wang:2025nhf, Ray:2026uur, Sridhar:2025kvi}, in contrast to the claims of some strongly modeled analyses~\citep{Tong:2025wpz, Antonini:2025ilj}.

\section{Future Work}
\label{sec:conclusion}
The novel astrophysical trends in the BBH population identified by our data-driven analysis framework motivate the development of targeted parametrizations that can yield stronger~(less uncertain) quantitative conclusions and constraints on relative abundances of various formation sub-channels. Such an investigation is ongoing. Data-driven analysis of other four-dimensional subspaces of BBH parameters, such as $(m_1,q,\chi_{eff},z)$, is in progress to uncover new correlations with redshifts. We are also scaling up our models to five dimensions~(see Appendix~\ref{sec:app-scale} for a discussion on such scalability), to constrain correlations in the space of $(m_1,q,\chi_{eff}, \chi_p, z)$ and determine the evolution of relative abundances of various formation pathways across cosmic time independent of prior restrictions. With upcoming data releases by the LVK, new discoveries from the growing BBH catalog, and a coherent astrophysical interpretation of observed subpopulations are likely imminent. 
\section{Acknowledgements}
We are grateful to Soumendra Kishore Roy for a detailed internal review of the manuscript and suggestions that have led to several improvements. We thank Fulya K{\i}ro{\u{g}}lu, Michael Zevin, and Matthew Mould for insightful discussions and suggestions. A.R. was supported by the National Science Foundation~(NSF) award PHY-2207945. V.K. was supported by the Gordon and Betty Moore Foundation (grant awards GBMF8477 and GBMF12341), through a Guggenheim Fellowship, and the D.I. Linzer Distinguished University Professorship fund. We are grateful for the computational resources provided by the LIGO Laboratory and supported by National Science Foundation Grants PHY-0757058 and PHY-0823459. This material is based upon work supported by NSF’s LIGO Laboratory, which is a major facility fully funded by the National Science Foundation. This research has made use of data obtained from the gravitational Wave Open Science Center (gwosc.org), a service of LIGO Laboratory, the LIGO scientific Collaboration, the Virgo Collaboration, and KAGRA. We gratefully acknowledge the support of the NSF-Simons AI-Institute for the Sky (SkAI) via grants NSF AST-2421845 and Simons Foundation MPS-AI-00010513.

We acknowledge the use of Large Language Models, specifically Anthropic's Claude (Fable-5) for help with transferring a CPU-based \texttt{Pymc} implementation of \texttt{gppop} to a multi-GPU-based \texttt{Pyro} implementation, and OpenAI's ChatGPT~(5.0) for generating plotting code and Overleaf tables. The analysis results, interpretation, and manuscript text are the sole responsibility of the authors.
\newpage
\appendix
\section{Model systematics}
\label{sec:app-sys}
As mentioned in the main text, our highly model-agnostic reconstruction of the BBH population can be sunject to additional systematics due to remaining restrictions. The effects of some of these on the results presented in the main text are easily probed while others require further investigation. We discuss these systematics below.

As a four-dimensional model with fixed $\kappa$, one can expect potential biases in the reconstructed distributions depending on the nature of the underlying correlations between $z$ and the modeled parameters provided the wrong values of $\kappa$ is chosen. Similarly, one can expect changes in some results depending on binning choices. Both of these can be investigated by comparing the inferred population corresponding to alternative binning choices and/or $\kappa$ values. We show in appendix~\ref{sec:app-varry-bink} that our results are fully robust against these systematics.

Due to their construction from component spins and mass ratios, parts of the $\chi_{eff}-\chi_p$ plane modeled in our analysis do not correspond to physical BBH systems. The non-trivial region of unphysicality should be imposed by setting the corresponding rate-densities to zero a priori, in contrast to our~(and alternate) current implementations~\citep{LIGOScientific:2026ctl}. However, due to a lack of data even in the physical regions neighbouring the unphysical ones, we do not expect these marginally non-zero rates to significantly affect our conclusions, which are based on the densities in the physical, data-reach regions.

Finally, as a smoothing prior, the choice of the covariance kernel of the GP might affect whether or not very sharp features in regions of sparse data are recovered by our model. The GP-prior prevents over-fitting for high-resolution models by correlating bins smaller than the inferred length scale~\citep{Ray:2023upk}. The inferred length scale therefore marks the smallest population feature in the corresponding BBH parameter that can be reconstructed from the given dataset. Whether or not this scale is affected by the functional form of the kernel requires further investigation. However, previous studies have shown that with enough observations, the likelihood dominates the inference and sharp features such as mass-gaps~\citep{Mandel:2016prl, Fishbach:2019ckx} and complex underlying correlations~\citep{Ray:2024hos} can be recovered successfully with the exponential quadratic kernel used in this work.
\section{Additional Results}
\label{sec:app-add}
\begin{figure*}
    \centering
    \includegraphics[width=0.98\textwidth]{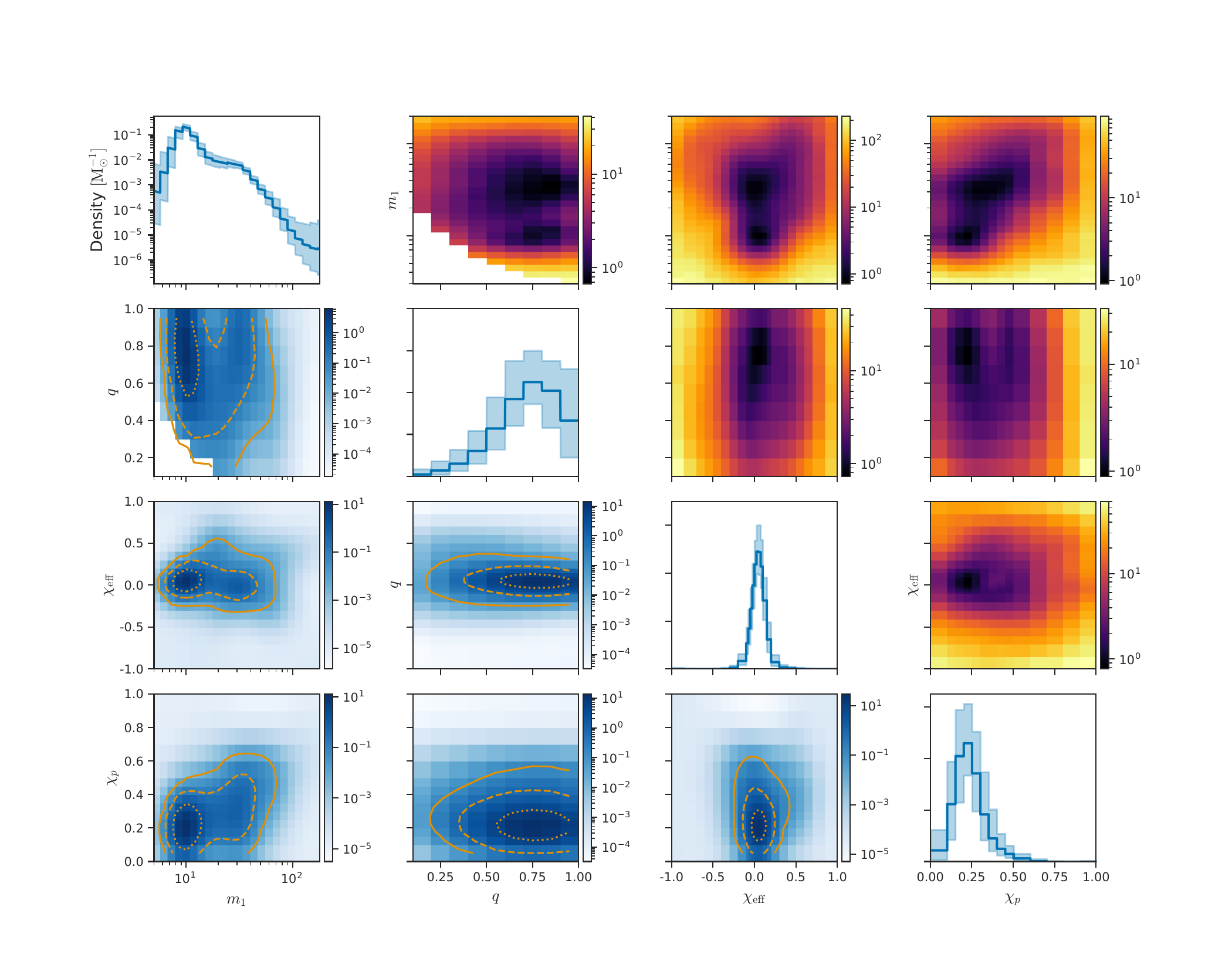}
    \caption{\label{fig:marginals-uc} Same as Figure~\ref{fig:marginals} but with the fractional uncertainties of the two-dimensional distributions presented on the upper-diagonal pannels.}
\end{figure*}

\begin{figure*}
    \centering
    \includegraphics[width=0.24\textwidth]{chieffchip_low.pdf}
    \includegraphics[width=0.24\textwidth]{chieffchip_20.pdf}
    \includegraphics[width=0.24\textwidth]{chieffchip_med.pdf}
    \includegraphics[width=0.24\textwidth]{chieffchip_high.pdf}
    \includegraphics[width=0.24\textwidth]{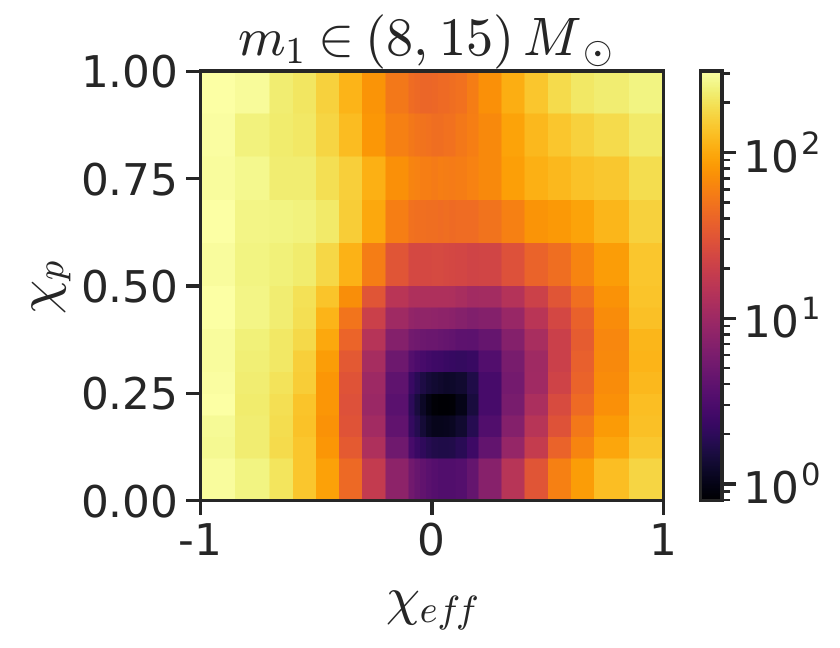}
    \includegraphics[width=0.24\textwidth]{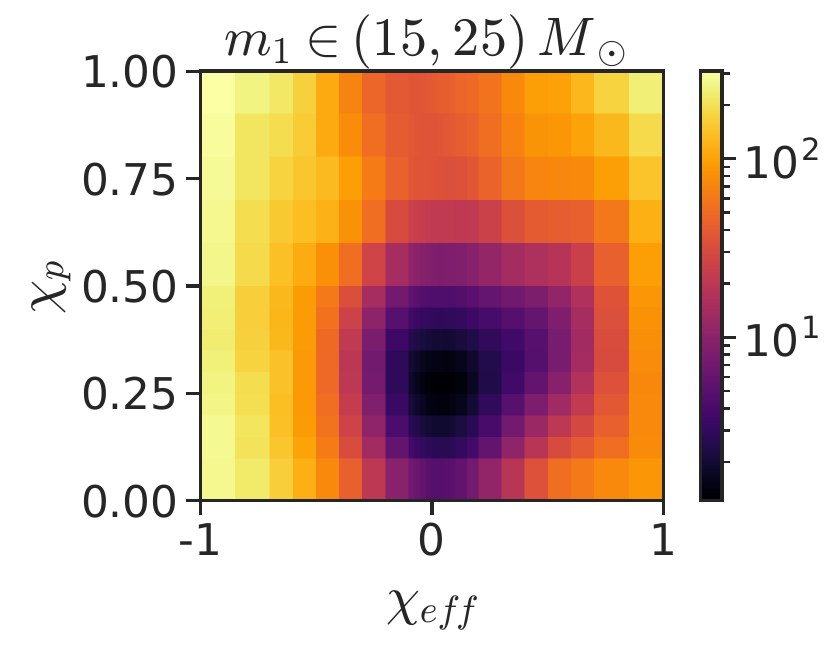}
    \includegraphics[width=0.24\textwidth]{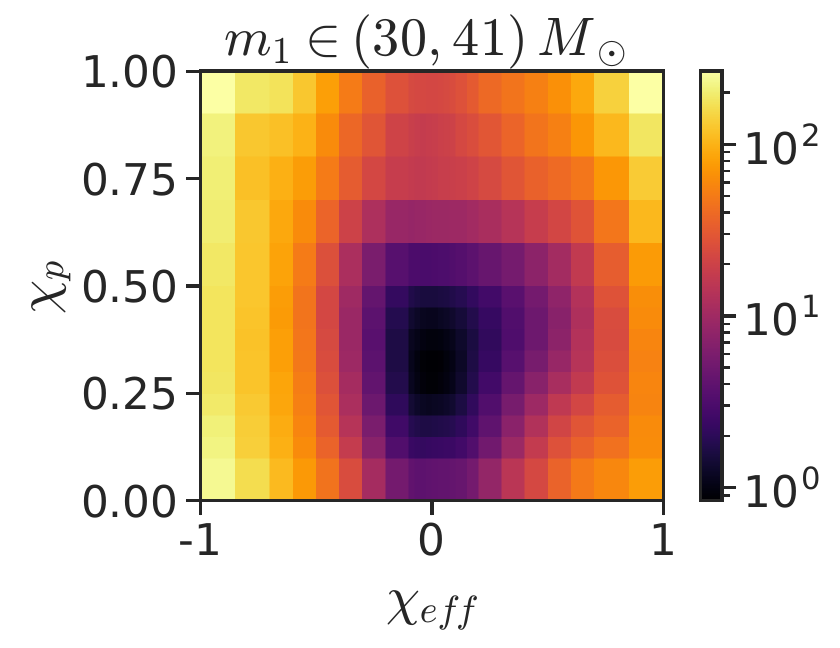}
    \includegraphics[width=0.24\textwidth]{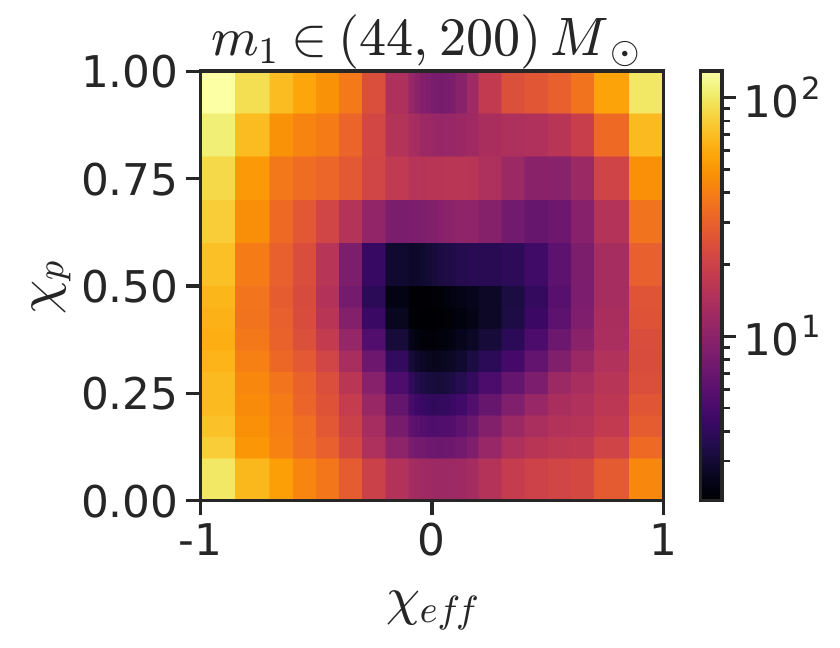}
    \includegraphics[width=0.24\textwidth]{qchieff_low.pdf}
    \includegraphics[width=0.24\textwidth]{qchieff_20.pdf}
    \includegraphics[width=0.24\textwidth]{qchieff_med.pdf}
    \includegraphics[width=0.24\textwidth]{qchieff_high.pdf}
    \includegraphics[width=0.24\textwidth]{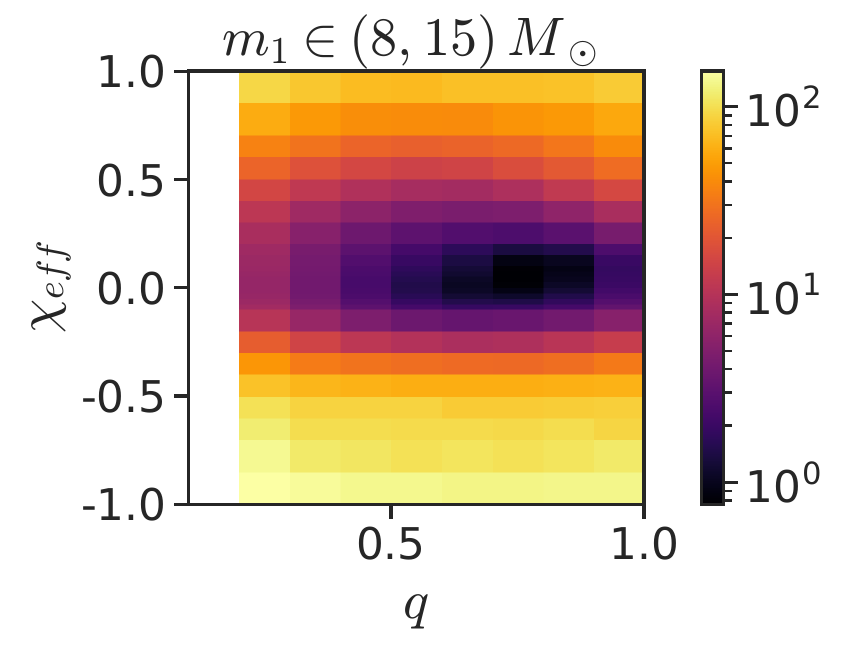}
    \includegraphics[width=0.24\textwidth]{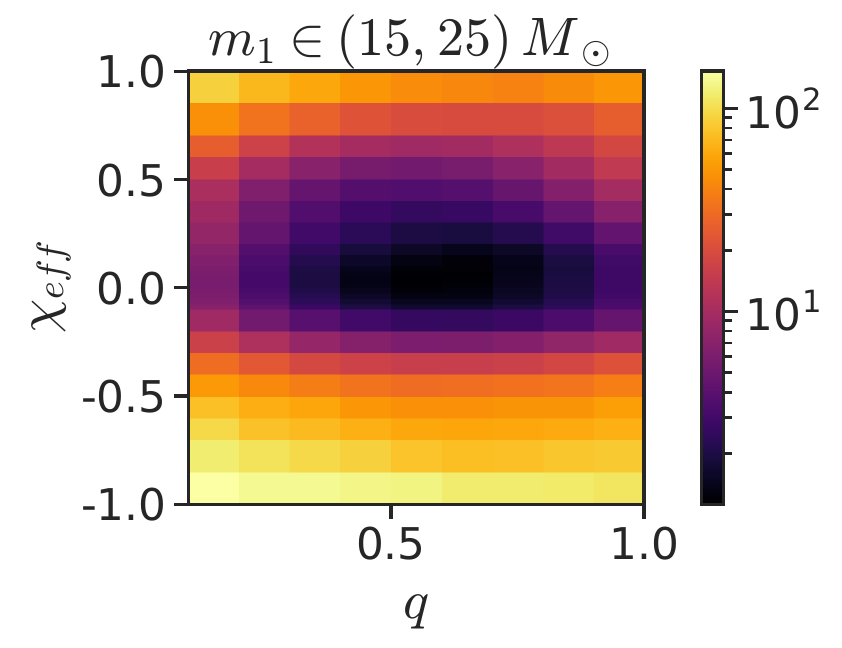}
    \includegraphics[width=0.24\textwidth]{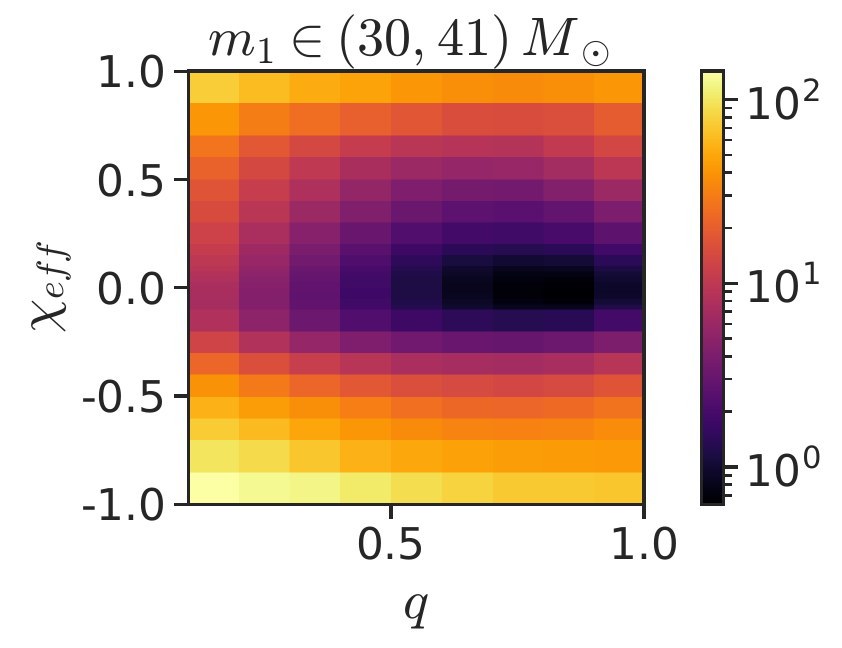}
    \includegraphics[width=0.24\textwidth]{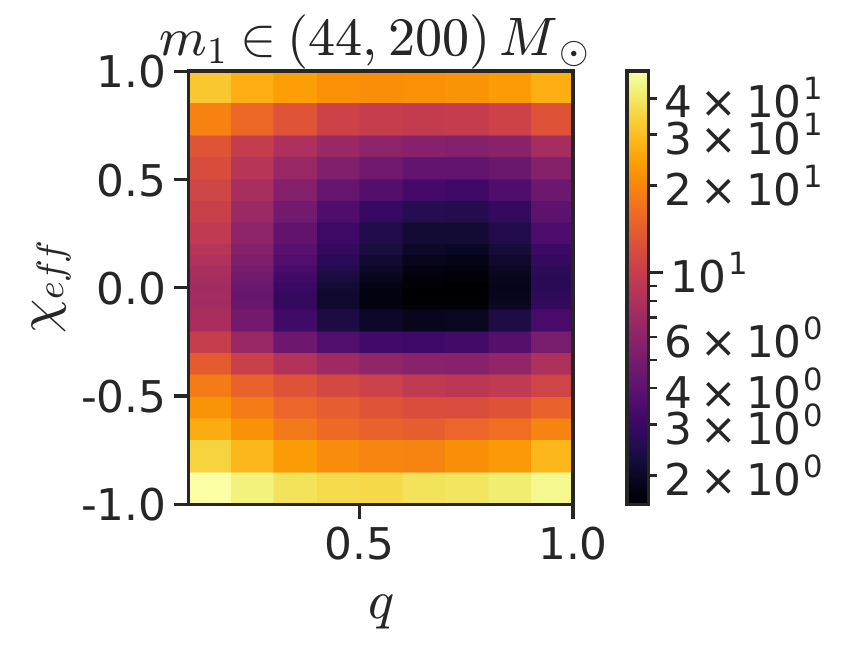}
    \includegraphics[width=0.24\textwidth]{qchip_low.pdf}
    \includegraphics[width=0.24\textwidth]{qchip_20.pdf}
    \includegraphics[width=0.24\textwidth]{qchip_med.pdf}
    \includegraphics[width=0.24\textwidth]{qchip_high.pdf}
    \includegraphics[width=0.24\textwidth]{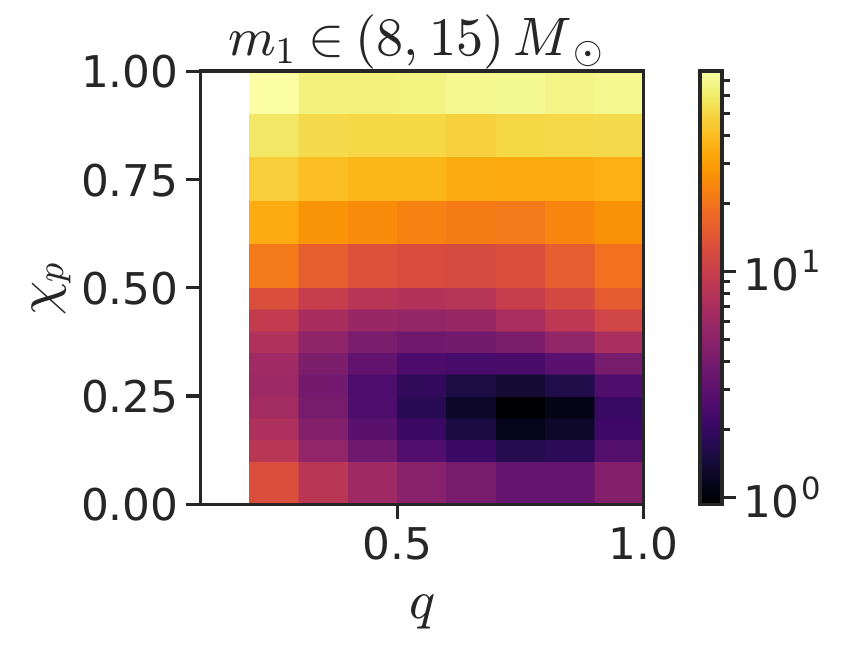}
    \includegraphics[width=0.24\textwidth]{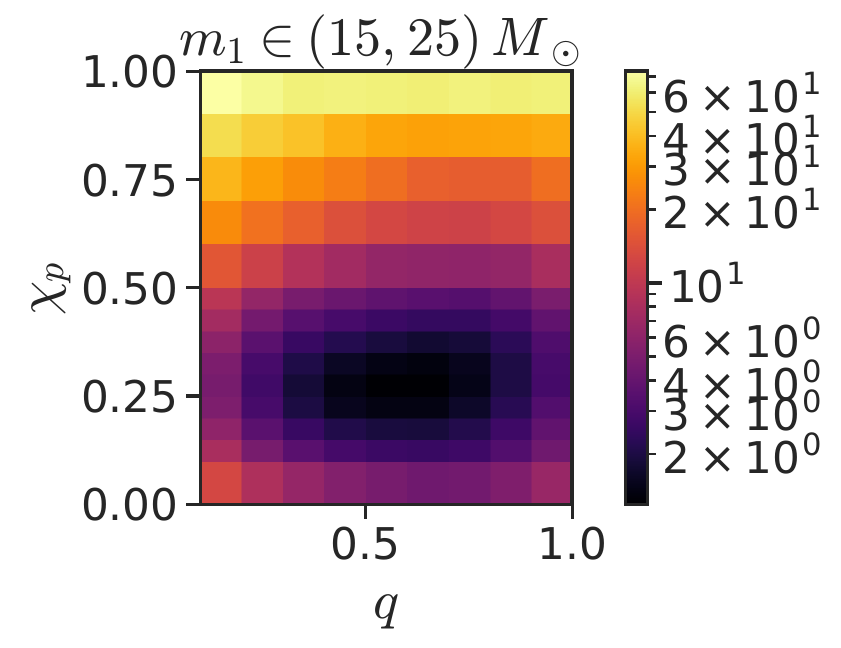}
    \includegraphics[width=0.24\textwidth]{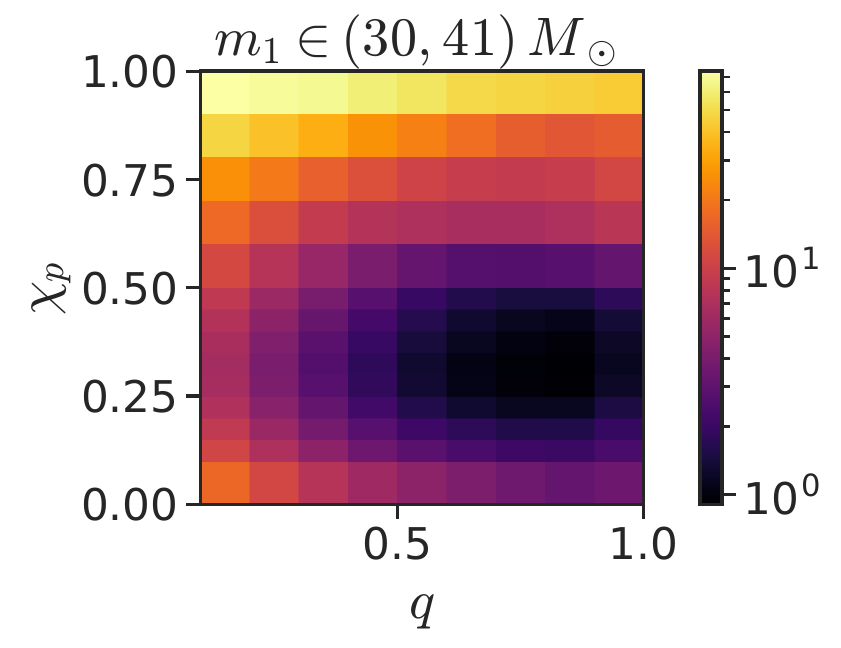}
    \includegraphics[width=0.24\textwidth]{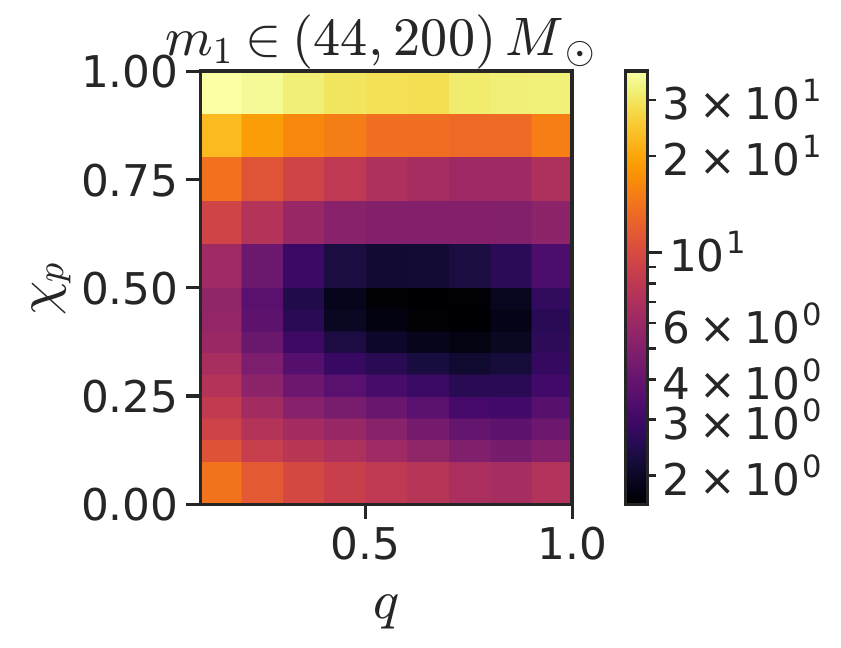}
    \caption{\label{fig:metrics}Posterior median~(\textit{first, third, and fifth row}) and fractional uncertainties (\textit{second, fourth, and sixth row}) of the two-dimensional distributions in different mass and mass ratio ranges.}
\end{figure*}

\begin{figure*}[t!]
    \centering
    \includegraphics[width=0.24\textwidth]{chieffchip_high.pdf}
    \includegraphics[width=0.24\textwidth]{chieffchip_44-55.pdf}
    \includegraphics[width=0.24\textwidth]{chieffchip_55-65.pdf}
    \includegraphics[width=0.24\textwidth]{chieffchip_65-200.pdf}
    \includegraphics[width=0.24\textwidth]{chieffchip_high_uc.pdf}
    \includegraphics[width=0.24\textwidth]{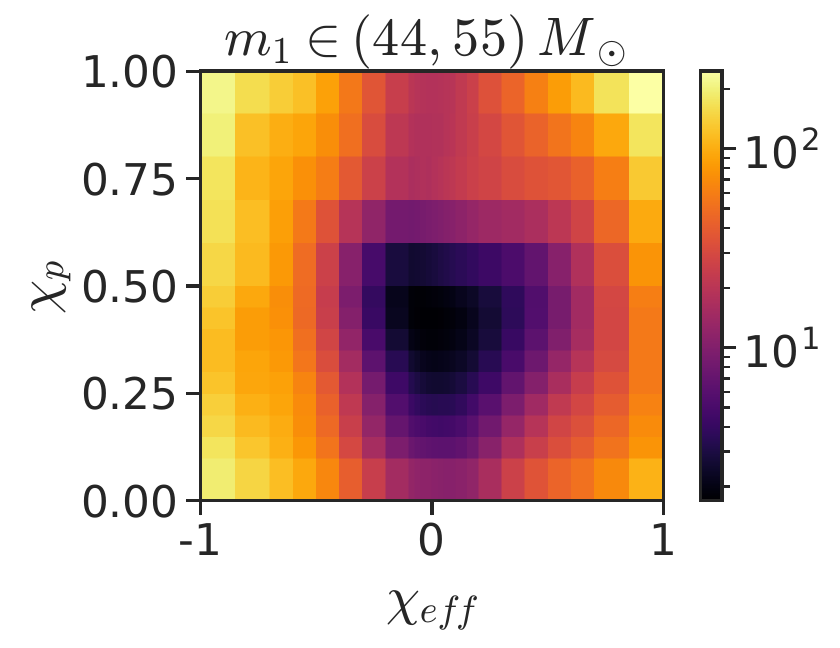}
    \includegraphics[width=0.24\textwidth]{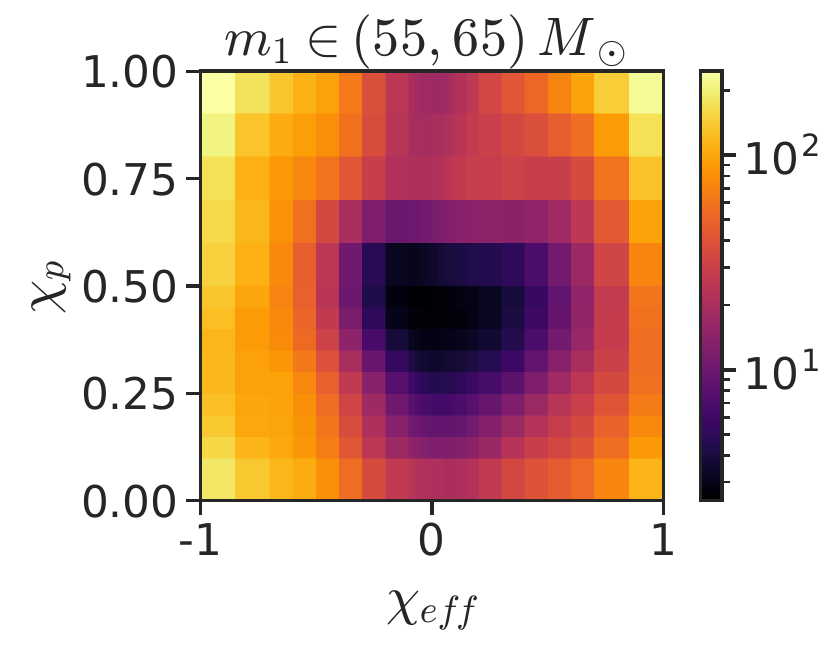}
    \includegraphics[width=0.24\textwidth]{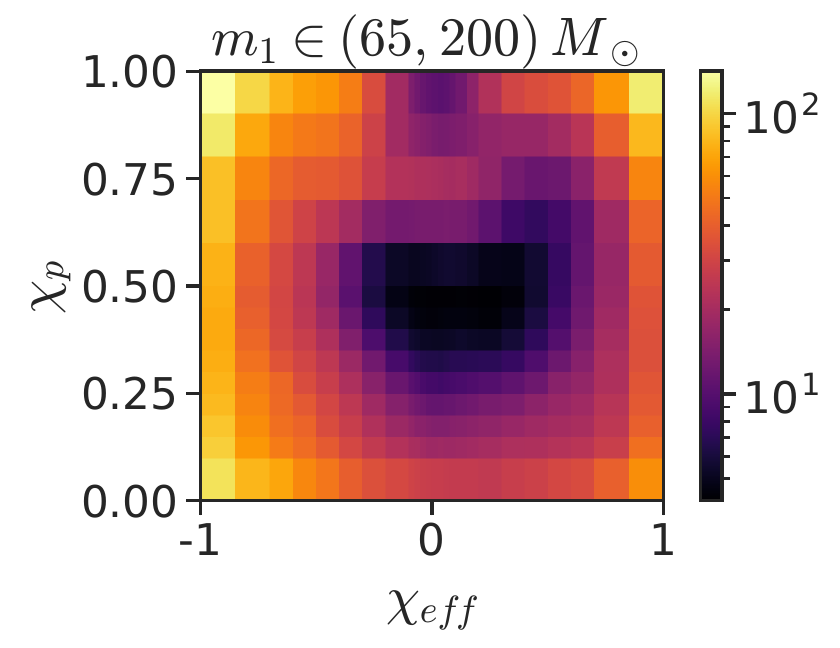}
    \caption{Posterior median (\textit{top}) and fractional uncertainty (\textit{bottom})of the $\chi_{eff}-\chi_p$ distribution in specific mass ranges above $44M_{\odot}$, \label{fig:trans-2Db-uc}}
\end{figure*}

\begin{figure*}
    \centering
    \includegraphics[width=0.24\textwidth]{chieffchip_low_lowq.pdf}
    \includegraphics[width=0.24\textwidth]{chieffchip_20_lowq.pdf}
    \includegraphics[width=0.24\textwidth]{chieffchip_med_lowq.pdf}
    \includegraphics[width=0.24\textwidth]{chieffchip_high_lowq.pdf}
    \includegraphics[width=0.24\textwidth]{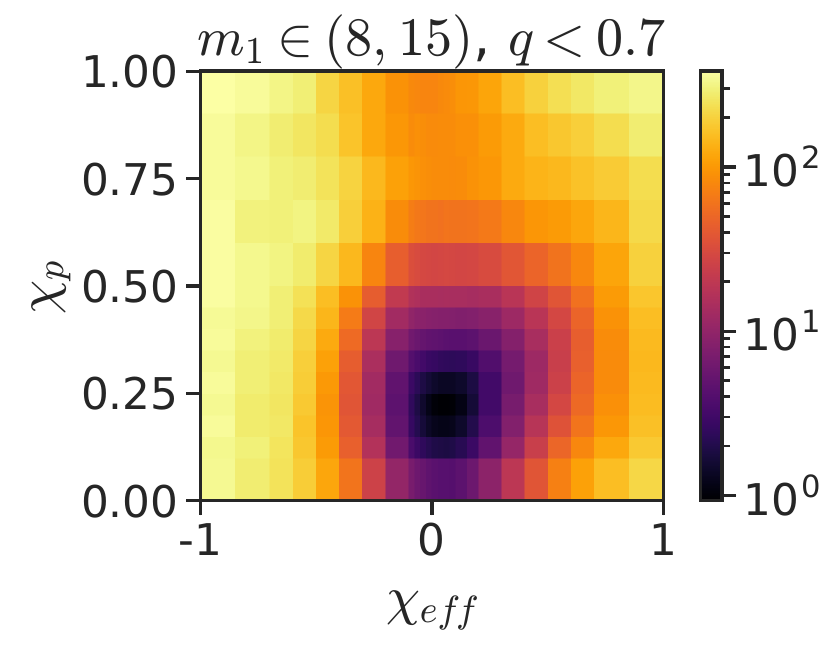}
    \includegraphics[width=0.24\textwidth]{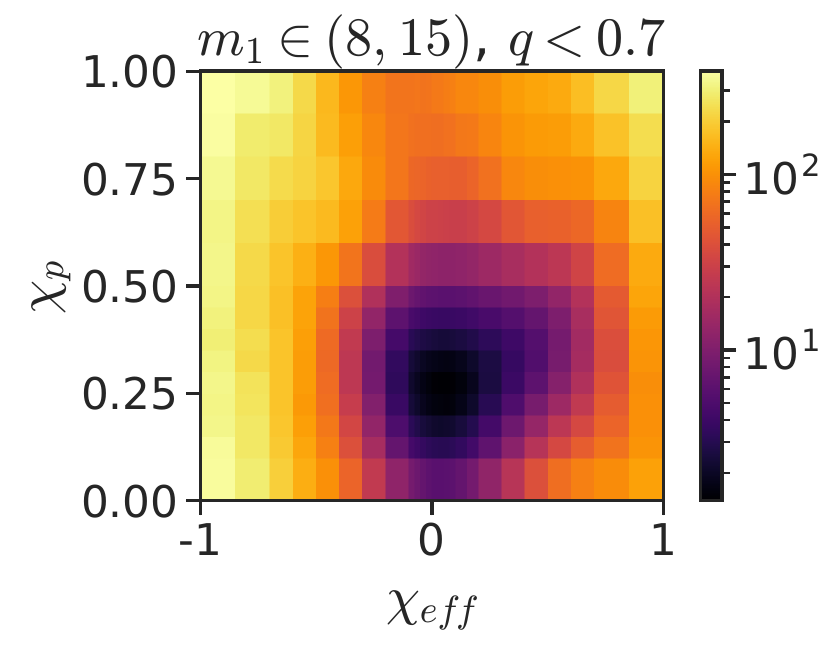}
    \includegraphics[width=0.24\textwidth]{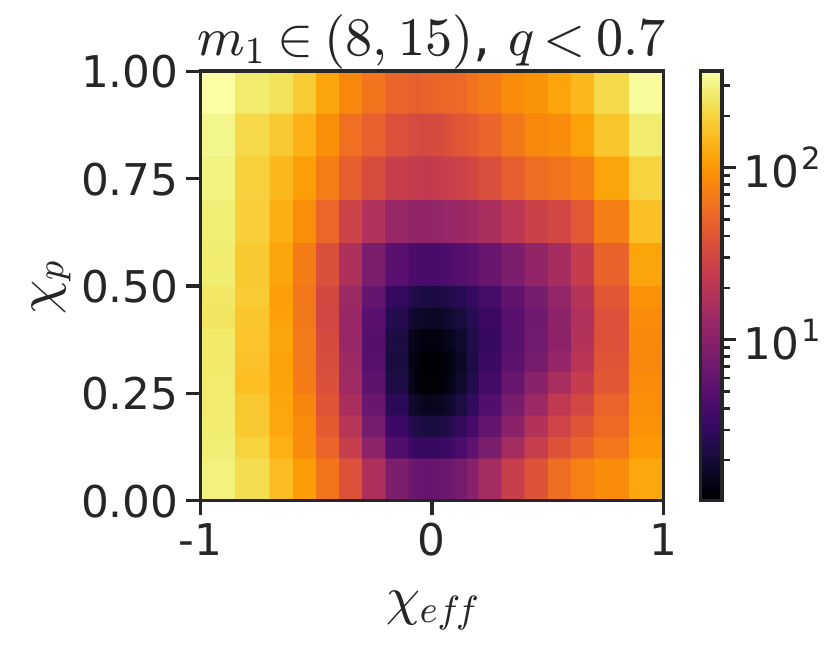}
    \includegraphics[width=0.24\textwidth]{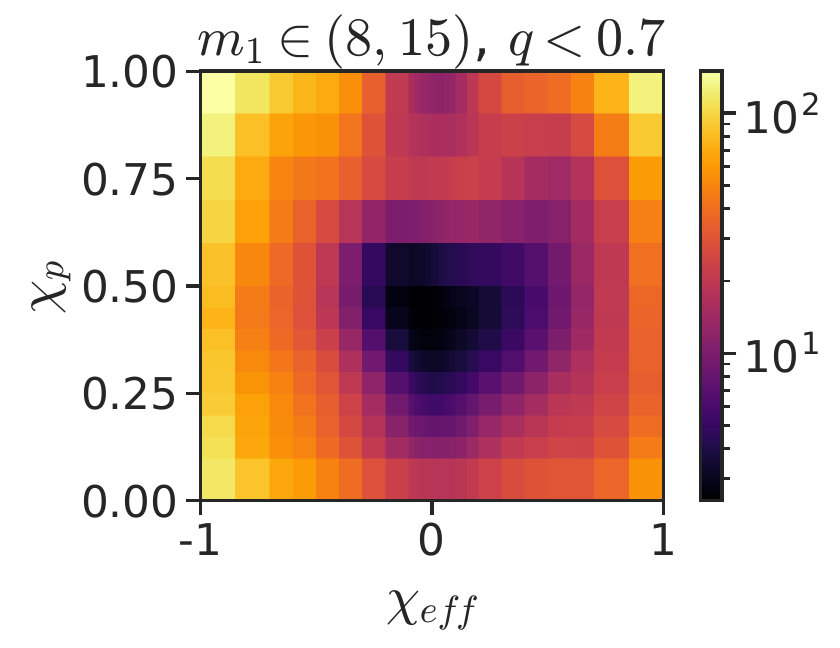}
    \includegraphics[width=0.24\textwidth]{chieffchip_low_highq.pdf}
    \includegraphics[width=0.24\textwidth]{chieffchip_20_highq.pdf}
    \includegraphics[width=0.24\textwidth]{chieffchip_med_highq.pdf}
    \includegraphics[width=0.24\textwidth]{chieffchip_high_highq.pdf}
    \includegraphics[width=0.24\textwidth]{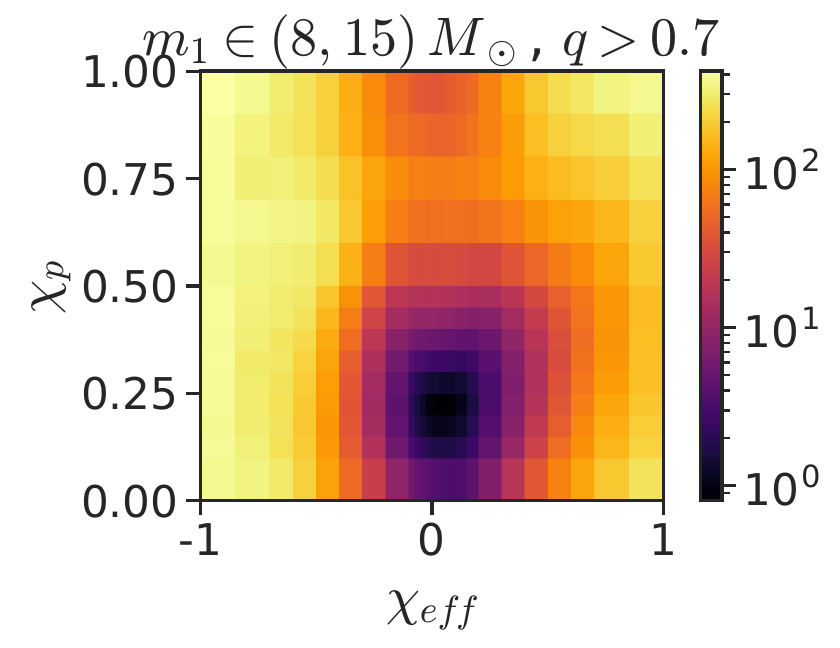}
    \includegraphics[width=0.24\textwidth]{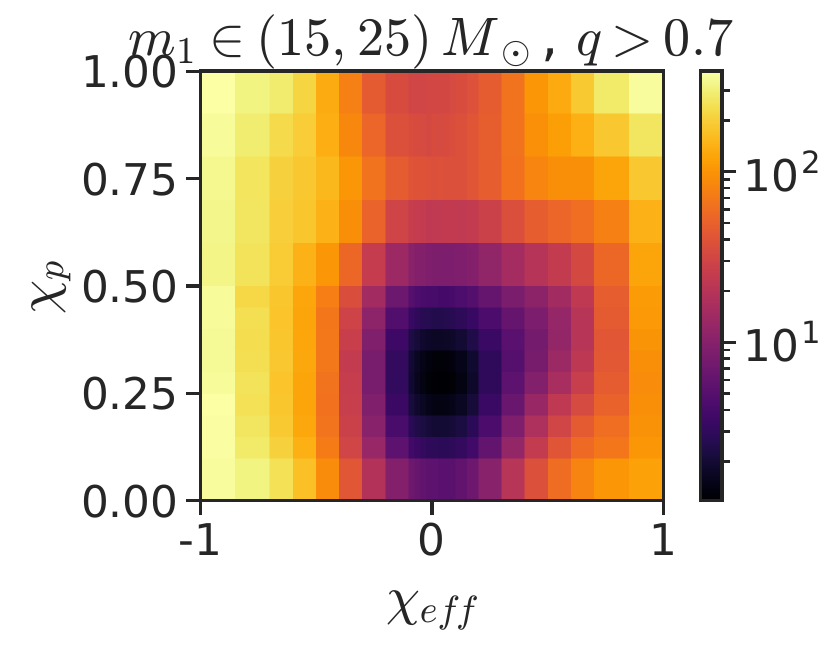}
    \includegraphics[width=0.24\textwidth]{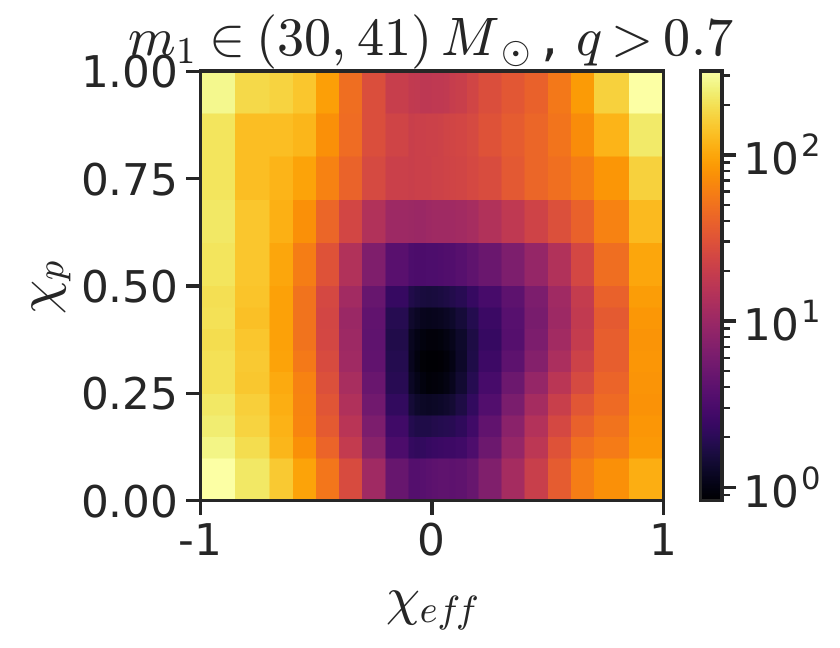}
    \includegraphics[width=0.24\textwidth]{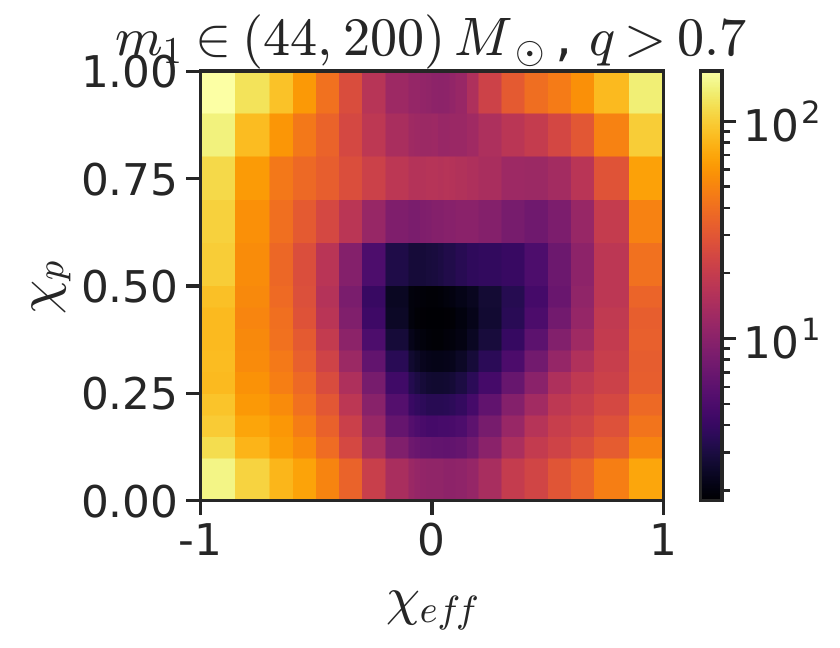}
    
    \caption{Posterior median~(\textit{first and third row}) and fractional uncertainties (\textit{second and fourth row}) of the $\chi_{eff}-\chi_p$ distributions in different mass and mass ratio ranges.\label{fig:trans-3D-uc}}
\end{figure*}

In this section, we present additional results to quantify the significance of various correlations identified in the posterior medians of the two-dimensional distributions shown in the main text. In particular, for each two-dimensional distribution whose posterior medians were presented in the main text, we present here the fractional uncertainties ($90\%$ credible interval scaled by the median) and compare them with the median, in Figures~\ref{fig:marginals-uc},~\ref{fig:metrics},~\ref{fig:trans-2Db-uc}, and~\ref{fig:trans-3D-uc}. We find that the regions of parameter space that exhibit the newly identified correlations have small fractional uncertainty which indicates that the corresponding astrophysical trends are statistically significant. However, as data-driven reconstructions, our conclusions are expected to be more uncertain than strongly modeled parametrizations. The goal is to design such parametrizations guided by the significant trends identified in this work which is an ongoing investigation.
% \begin{figure*}
%     \centering
%     \includegraphics[width=0.24\linewidth]{example-image-a}
%     \includegraphics[width=0.24\linewidth]{example-image-a}
%     \includegraphics[width=0.24\linewidth]{example-image-a}
%     \includegraphics[width=0.24\linewidth]{example-image-a}
%     \includegraphics[width=0.24\linewidth]{example-image-a}
%     \includegraphics[width=0.24\linewidth]{example-image-a}
%     % \includegraphics[width=0.24\linewidth]{muchieff_q_given_m1.pdf}
%     % \includegraphics[width=0.24\linewidth]{sigchieff_q_given_m1.pdf}
%     \includegraphics[width=0.24\linewidth]{example-image-a}
%     \includegraphics[width=0.24\linewidth]{example-image-a}
%     \includegraphics[width=0.24\linewidth]{example-image-a}
%     \includegraphics[width=0.24\linewidth]{example-image-a}
%     \includegraphics[width=0.24\linewidth]{example-image-a}
%     \includegraphics[width=0.24\linewidth]{example-image-a}
%     \caption{\label{fig:metrics}Credible intervals of the mean and standard deviations of one BBH parameters as a function of others.\textcolor{red}{Supplemental results are waiting on a few reruns and will be updated in v2}}
% \end{figure*}

\section{Varying Bin Resolution and Values of  $\kappa$}

\label{sec:app-varry-bink}
\begin{table*}
% \begin{adjustwidth}{0cm}{}
\centering
\begin{tabular}{cc}
\hline
\hline
Parameter & Bin edges \\
\hline
$m_1 (M_{\odot})$           & \hspace{1em} log-uniform(3, 200, 25)                                                                              \\

\hline
$\chi_{\rm eff}$ & \hspace{1em} uniform (-1,1,15)                                                                 \\
\hline
$\chi_p$   & \hspace{1em} uniform (0,1,11)  \\
\hline
$q$   & \hspace{1em} uniform(0.11,1,10)\\
\hline
\end{tabular}
% \end{adjustwidth}
\caption{The bin edges used for Model Z.}\label{tab: Bin choices Z}
\end{table*} 
\begin{figure*}
    \centering
    \includegraphics[width=0.32\textwidth]{pq_given_m1.pdf}
    \includegraphics[width=0.32\textwidth]{pchieff_given_m1.pdf}
    \includegraphics[width=0.32\textwidth]{pchip_given_m1.pdf}
    \includegraphics[width=0.32\textwidth]{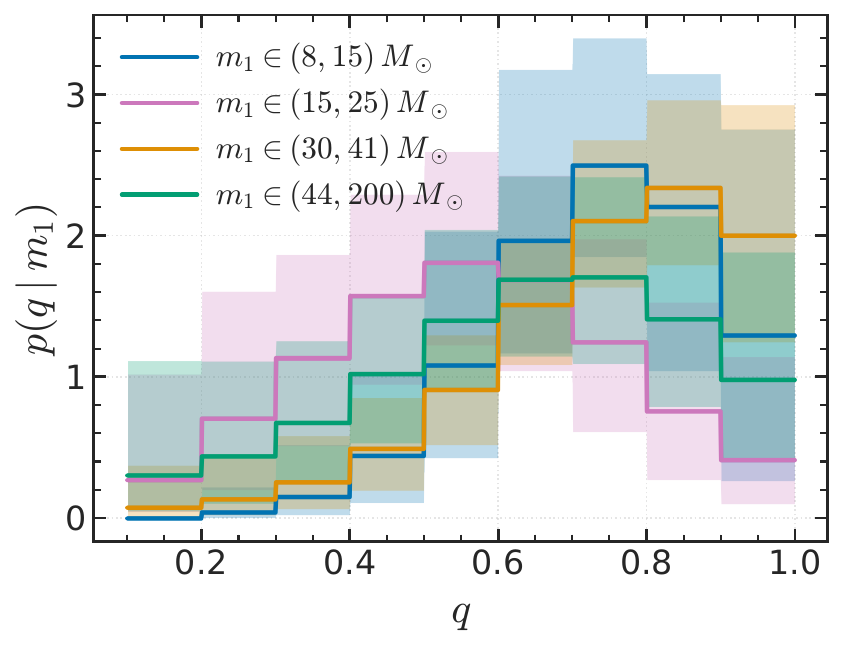}
    \includegraphics[width=0.32\textwidth]{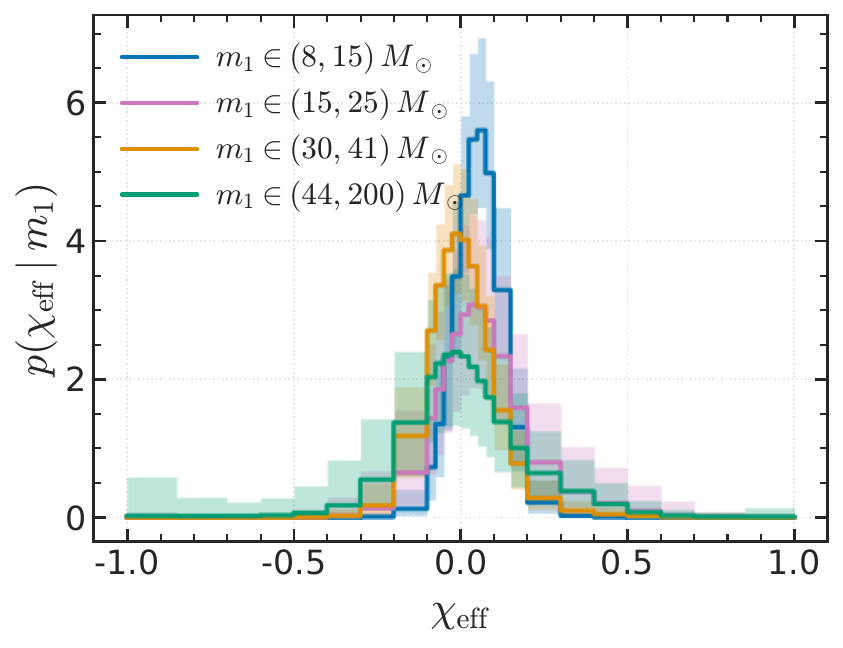}
    \includegraphics[width=0.32\textwidth]{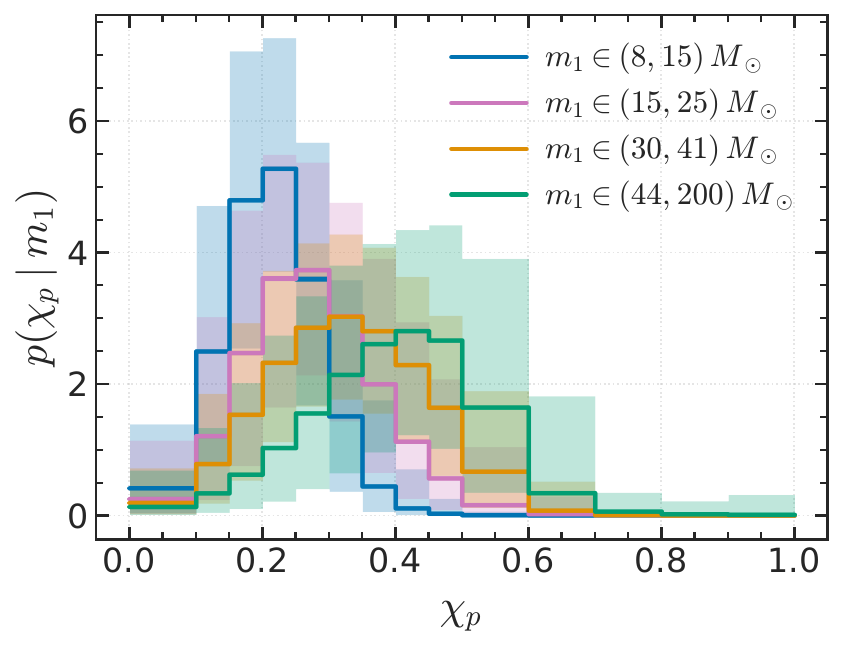}
    \includegraphics[width=0.32\textwidth]{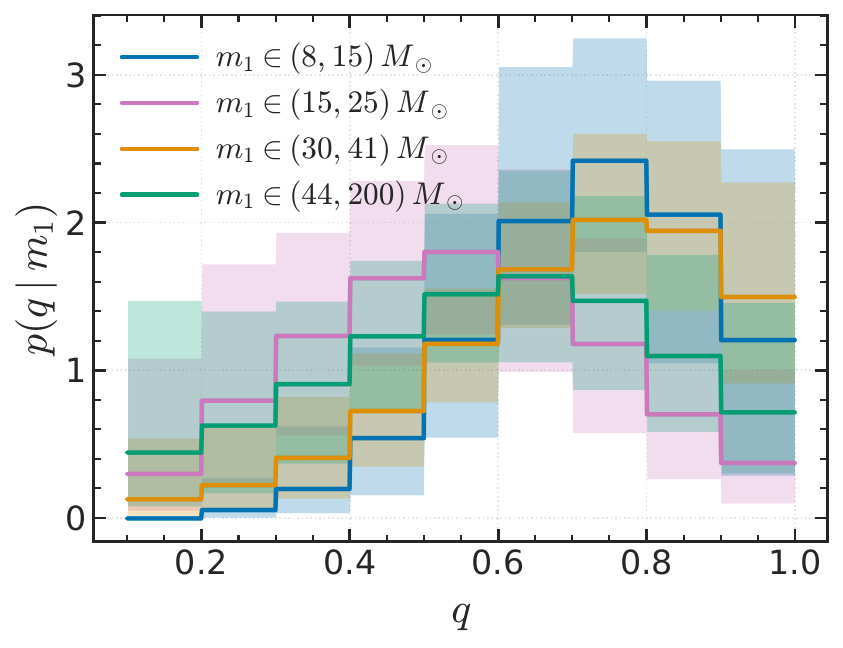}
    \includegraphics[width=0.32\textwidth]{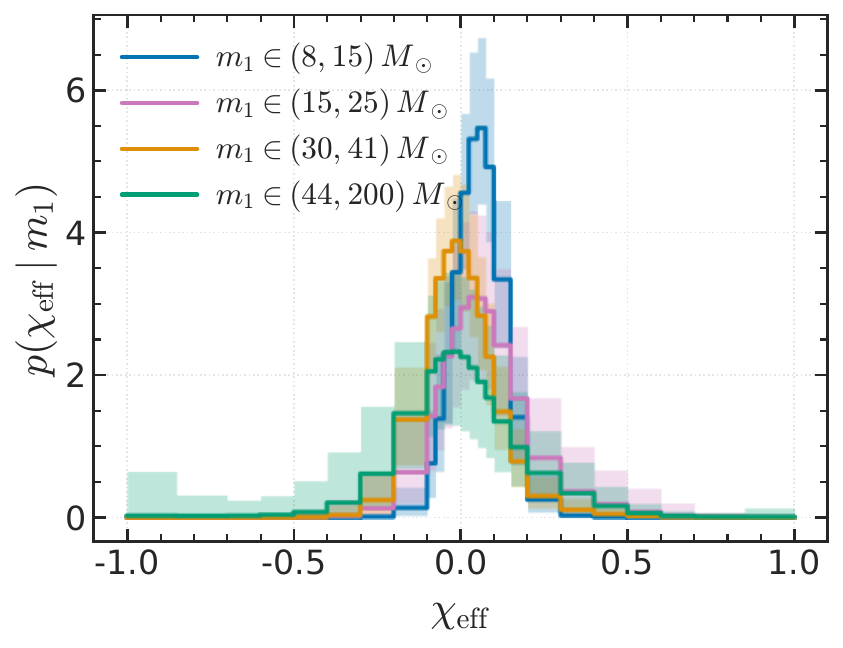}
    \includegraphics[width=0.32\textwidth]{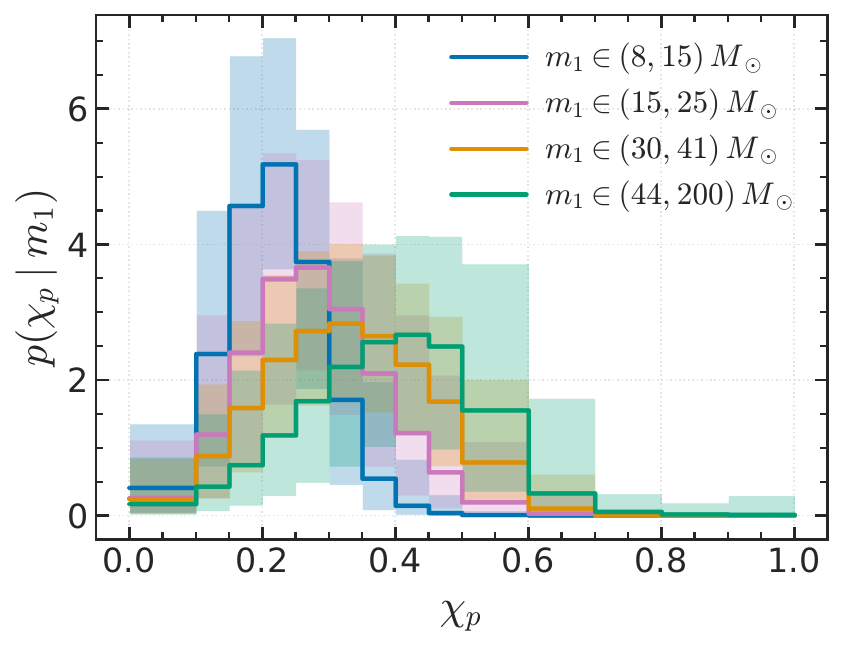}
    \includegraphics[width=0.32\textwidth]{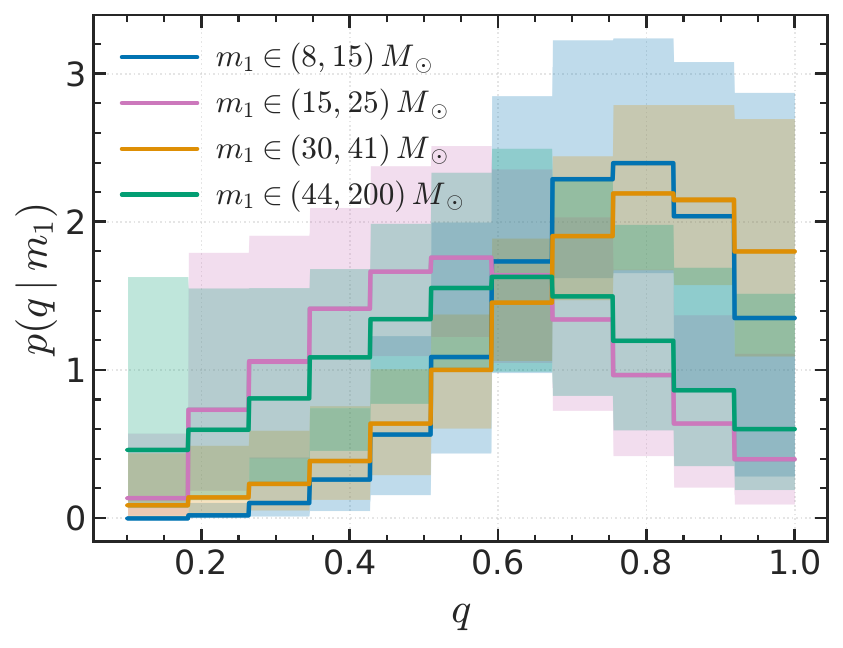}
    \includegraphics[width=0.32\textwidth]{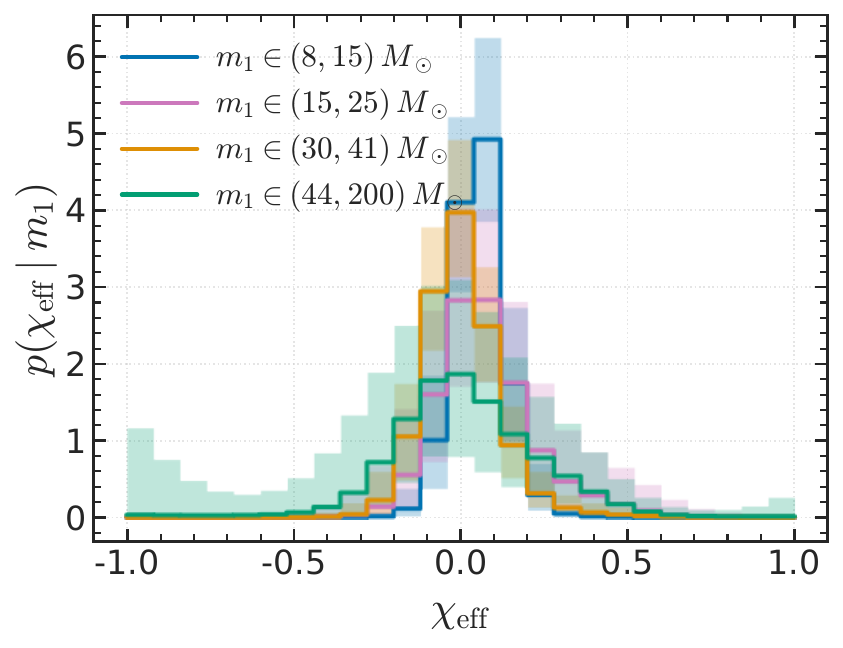}
    \includegraphics[width=0.32\textwidth]{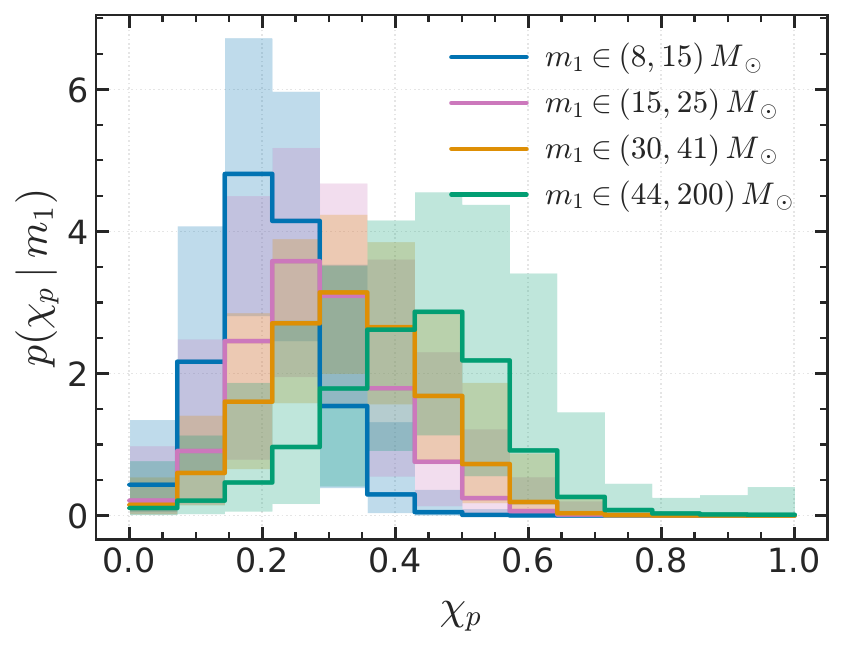}
    \caption{One dimensional distributions conditioned on primary mass for the default binning choice (\textit{row 1}), and models X,Y,Z (\textit{rows 2,3,4}) \label{fig:trans-1Dmc}}
\end{figure*}
In this section, we test the robustness of our results against variations in binning choices and $\kappa$ values. In addition to the default binning scheme and $\kappa$ values, we consider three additional models. Models X, and Y use the same binning choices as the default one but use $\kappa=1.9$ and $\kappa=3.9$, respectively. For Model Z, we choose the same $\kappa=2.7$ as the default model but use alternate binnings summarized in Table~\ref{tab: Bin choices Z}.

\begin{figure*}
    \centering
    \includegraphics[width=0.24\linewidth]{chieffchip_high.pdf}
    \includegraphics[width=0.24\linewidth]{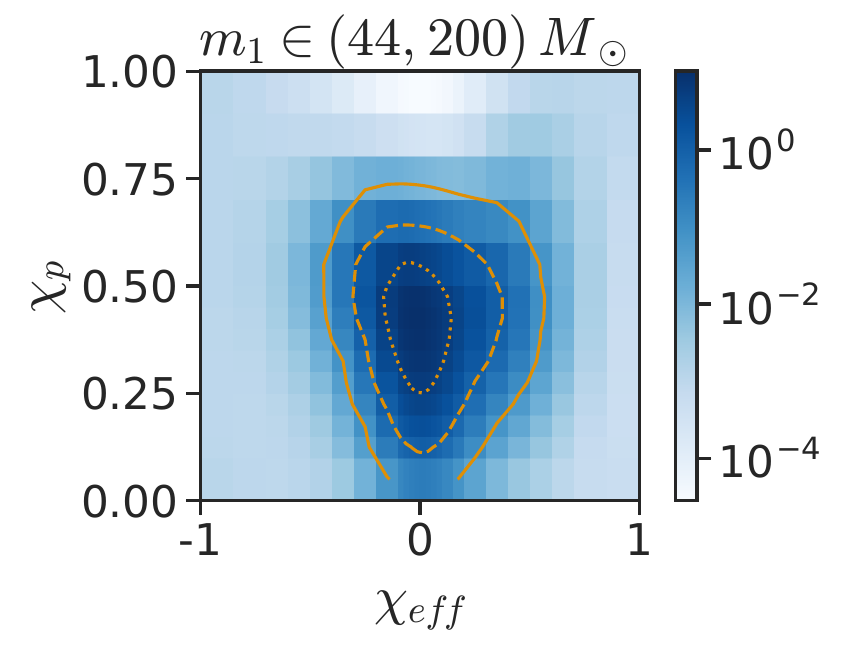}
    \includegraphics[width=0.24\linewidth]{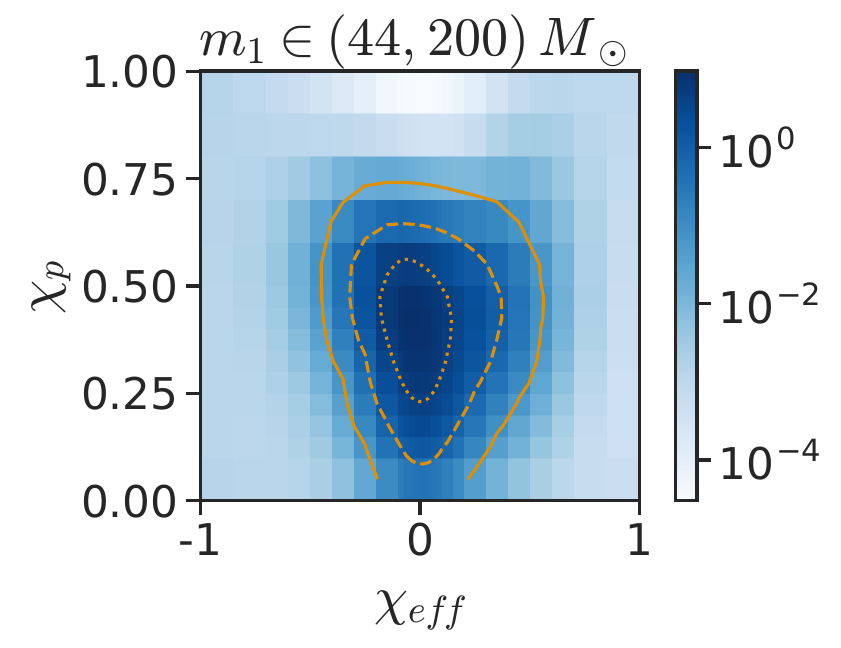}
    \includegraphics[width=0.24\linewidth]{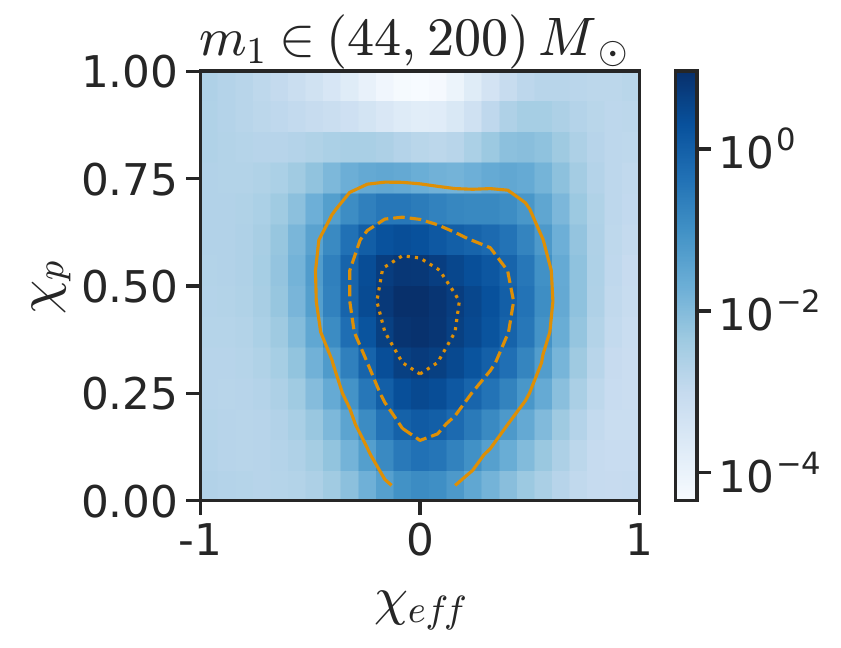}
    \includegraphics[width=0.24\linewidth]{qchieff_20.pdf}
    \includegraphics[width=0.24\linewidth]{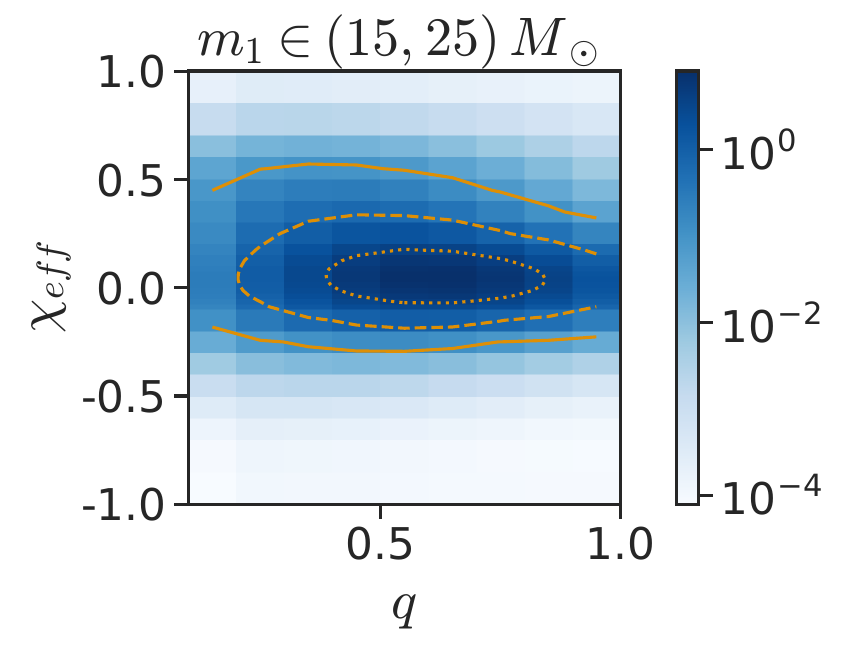}
    \includegraphics[width=0.24\linewidth]{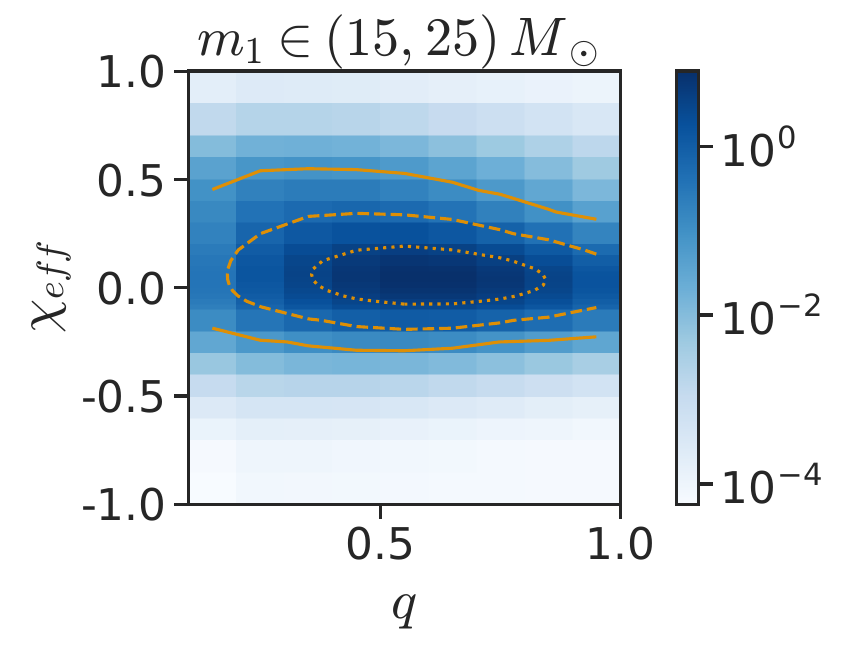}
    \includegraphics[width=0.24\linewidth]{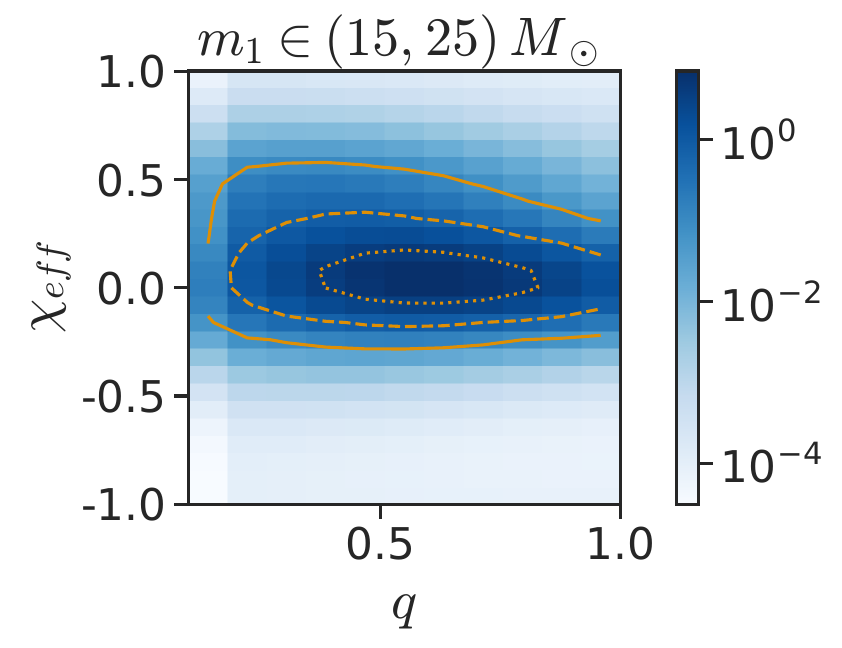}
    \includegraphics[width=0.24\linewidth]{qchieff_med.pdf}
    \includegraphics[width=0.24\linewidth]{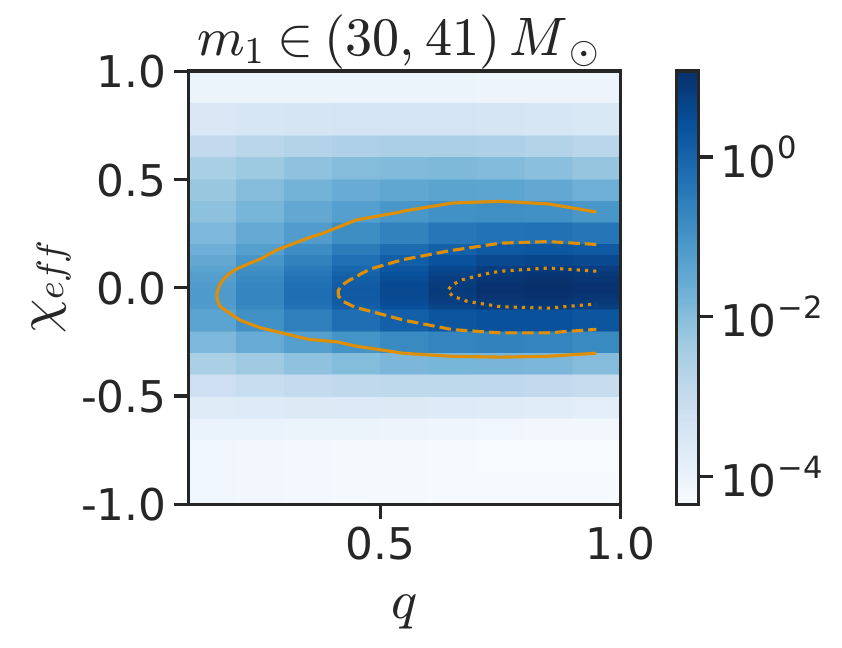}
    \includegraphics[width=0.24\linewidth]{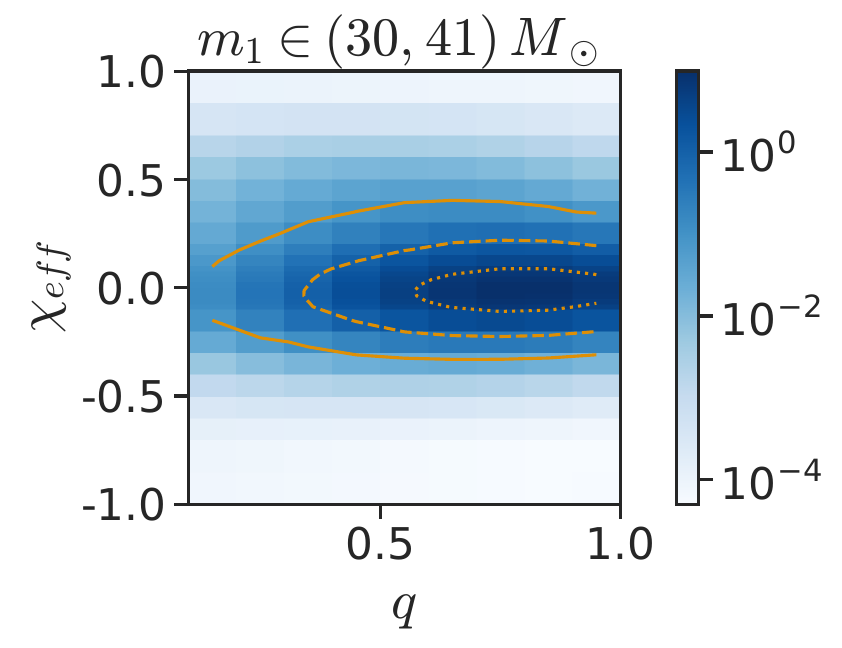}
    \includegraphics[width=0.24\linewidth]{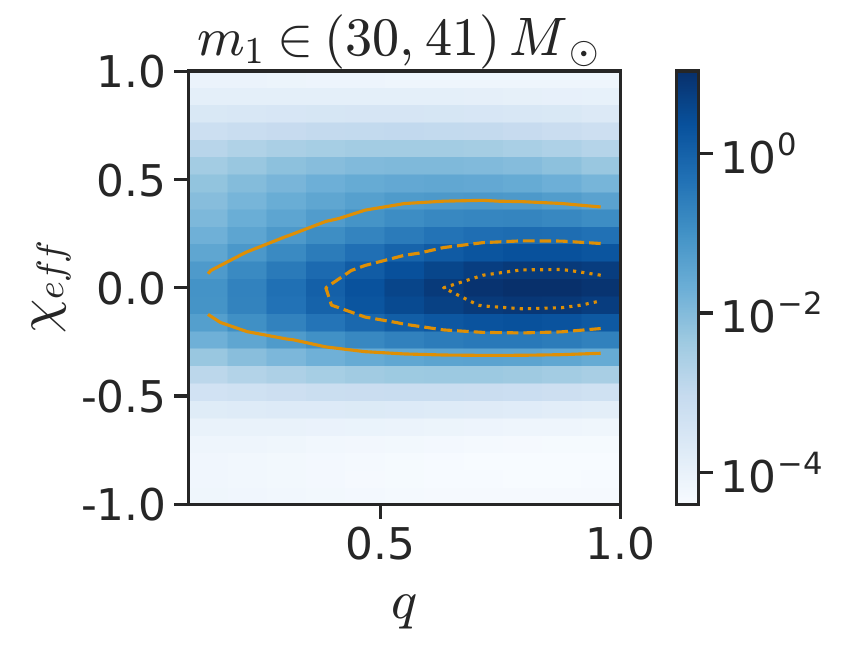}
    \includegraphics[width=0.24\linewidth]{qchip_med.pdf}
    \includegraphics[width=0.24\linewidth]{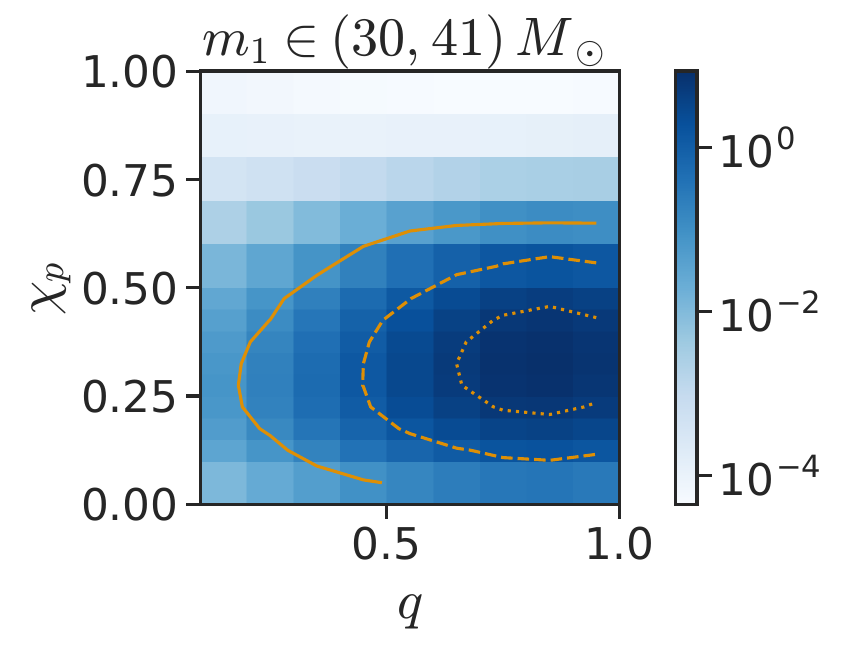}
    \includegraphics[width=0.24\linewidth]{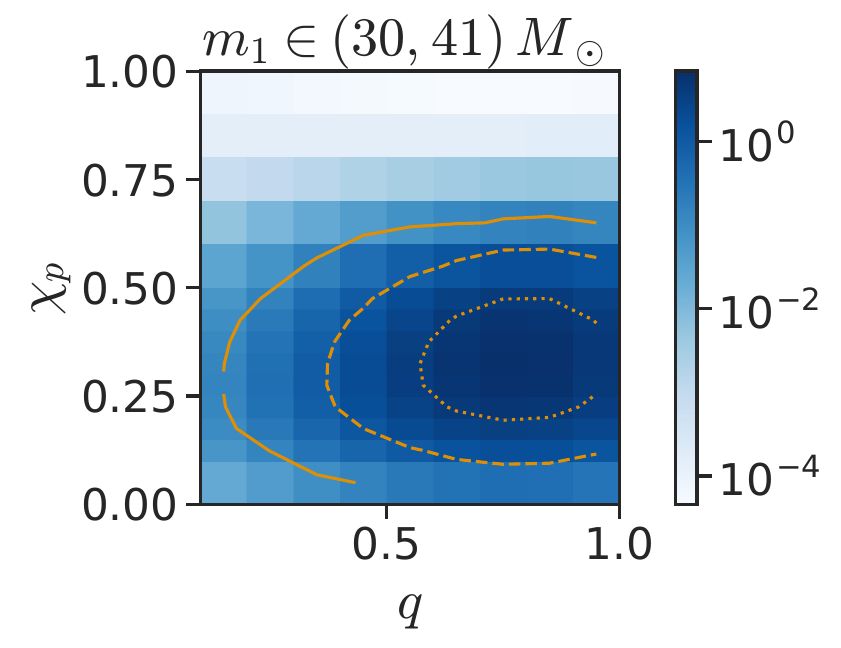}
    \includegraphics[width=0.24\linewidth]{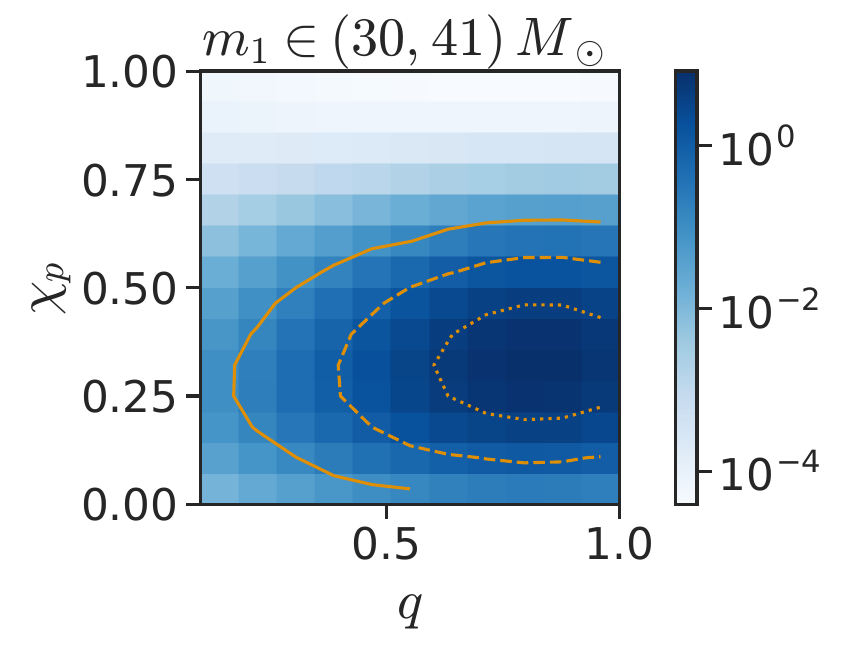}
    \caption{\label{fig:rob-borad}Robustness of the $\chi_{eff}-\chi_p$~(\textit{row 1}) broadening at high masses, the $\chi_{eff}-q$ broadening in the $15-25M_{\odot}$~(\textit{row 2}) and $30-40$~(row 3) mass ranges, and the $\chi_p-q$ broadening in the $30-40M_{\odot}$ range~(\textit{row 4}) against variations in binning and $\kappa$ values (\textit{from left to right models default, X, Y, and Z}).}
\end{figure*}

To avoid clutter, we do not reproduce every single plot presented in the main text for all three models. We instead show in Figures~\ref{fig:trans-1Dmc}, and~\ref{fig:rob-borad}, the key results, namely the conditional one-dimensional distributions in various mass ranges, the joint $\chi_{eff}-\chi_p$ broadening at high masses, the $q-\chi_{eff}$ distributions in the $15-20M_{\odot}$ and $30-40M_{\odot}$ mass ranges, and the $\chi_p-q$ broadening in the $30-40M_{\odot}$ mass range for all three models. We find that the new correlations presented in the main text are fully robust against variations in binning choices and $\kappa$ values. %We further compute the posterior on the JS divergences between the joint four-dimensional inferred populations of all 3 models and the default one, which are presented in Figure~\ref{fig:rob-js}. 

% \begin{figure*}
%     \centering
%     \includegraphics[width=0.98\linewidth]{example-image-a}
%     \caption{\label{fig:rob-js}JS divergences between the inferred four dimensional distributions corresponding to different binning choices and $\kappa$ values.}
% \end{figure*}
\section{Hyperposterior and Variance Penalty}
For our BGP model, the hierarchical likelihood takes the following form:
\begin{equation}
    \log p(\vec{d}|\vec{n}) =   -\sum_\gamma n^\gamma \left<VT\right>^\gamma + \sum_i \log \biggl(\sum_\gamma n^\gamma w^\gamma (d_i)\biggr),
\end{equation}
where, $\left<VT\right>^\gamma$ and $w^\gamma (d_i)$ are Monte Carlo estimators of the detectable hypervolume and event-specific posterior supports from the ith event, in the $\gamma$th bin respectively. The means~($\mu_{VT}^{\gamma},\mu_{w}^{\gamma}(d_i)$) and variances~($(\sigma_{VT}^{\gamma})^2,(\sigma_{w}^{\gamma}(d_i))^2$) of these estimators can be computed as follows:
\begin{equation}
\begin{aligned}
\mu_{w}^{\gamma}(d_i)
={}& \frac{1}{N_{samp}}
\sum_{j\sim p(\vec{\theta}\mid d_i)}
\frac{1}{
p_{\mathrm{PE}}\!\left(
m_{1,j}, q_j, \chi_{\mathrm{eff},j},
\chi_{p,j}, z_j
\right)
}
\frac{1}{m_{1,j}}
\left.\frac{dV}{dz}\right|_{z_j}
(1+z_j)^{\kappa-1}
\\
&~\quad\quad\quad\quad\quad\quad\times
\begin{cases}
1,
&
\left(
m_{1,j}, q_j, \chi_{\mathrm{eff},j},
\chi_{p,j}
\right)
\in \gamma^{\mathrm{th}}\text{ bin},
\\
0,
&
\text{o.w.}
\end{cases}
\end{aligned}
\end{equation}

\begin{equation}
\begin{aligned}
\mu_{VT}^{\gamma}
={}&\frac{1}{N_{draw}}
\sum_{j\sim p(\vec{\theta}\mid \text{det,draw})}
\frac{1}{
p_{\mathrm{draw}}\!\left(
m_{1,j}, q_j, \chi_{\mathrm{eff},j},
\chi_{p,j}, z_j
\right)
}
\frac{1}{m_{1,j}}
\left.\frac{dV}{dz}\right|_{z_j}
(1+z_j)^{\kappa-1}
\\
&~\quad\quad\quad\quad\quad\quad\quad\quad\times
\begin{cases}
1,
&
\left(
m_{1,j}, q_j, \chi_{\mathrm{eff},j},
\chi_{p,j}
\right)
\in \gamma^{\mathrm{th}}\text{ bin},
\\
0,
&
\text{o.w.}
\end{cases}
\end{aligned}
\end{equation}
\begin{equation}
\begin{aligned}
(\sigma_{w}^{\gamma}(d_i))^2
={}& -\frac{(\mu_{w}^{\gamma}(d_i))^2}{N_{samp}}+\frac{1}{N^2_{samp}}
\sum_{j\sim p(\vec{\theta}\mid d_i)}
\left[\frac{1}{
p_{\mathrm{PE}}\!\left(
m_{1,j}, q_j, \chi_{\mathrm{eff},j},
\chi_{p,j}, z_j
\right)
}
\frac{1}{m_{1,j}}
\left.\frac{dV}{dz}\right|_{z_j}
(1+z_j)^{\kappa-1}\right]^2
\\
&~\quad\quad\quad\quad\quad\quad\quad\quad\quad\quad\quad\quad\quad\quad\quad\times
\begin{cases}
1,
&
\left(
m_{1,j}, q_j, \chi_{\mathrm{eff},j},
\chi_{p,j}
\right)
\in \gamma^{\mathrm{th}}\text{ bin},
\\
0,
&
\text{o.w.}
\end{cases}
\end{aligned}
\end{equation}

\begin{equation}
\begin{aligned}
(\sigma_{VT}^{\gamma})^2
={}& -\frac{(\mu_{VT}^{\gamma})^2}{N_{draw}}+\frac{1}{N^2_{draw}}
\sum_{j\sim p(\vec{\theta}\mid \text{det, draw})}
\left[\frac{1}{
p_{\mathrm{draw}}\!\left(
m_{1,j}, q_j, \chi_{\mathrm{eff},j},
\chi_{p,j}, z_j
\right)
}
\frac{1}{m_{1,j}}
\left.\frac{dV}{dz}\right|_{z_j}
(1+z_j)^{\kappa-1}\right]^2
\\
&~\quad\quad\quad\quad\quad\quad\quad\quad\quad\quad\quad\quad\quad\quad\quad\quad\quad\times
\begin{cases}
1,
&
\left(
m_{1,j}, q_j, \chi_{\mathrm{eff},j},
\chi_{p,j}
\right)
\in \gamma^{\mathrm{th}}\text{ bin},
\\
0,
&
\text{o.w.}
\end{cases}
\end{aligned}
\end{equation}
where, $p(\vec{\theta}|d_i)$ is the single event parameter estimation posterior for the $ith$ event, samples of which are publicly released by the LVK, $p_{PE}$ is the population agnostic prior used in said PE marginalized over other parameters, $N_{samp}$ is the number of single-event posterior samples, $p(\vec{\theta}|\text{det,draw})$ represents the distribution whose samples are the ranked simulated signals found above the detection threshold used in event selection which are also publicly released by the LVK, $p_{draw}$ is the fiducial population from which the simulated signals were generated marginalized over other parameters, and $N_{draw}$ are the total number of simulations generated regardless of rank.

These means and variances are used to evaluate both the log-likelihood estimator and its variance due to Monte Carlo uncertainties~\citep{Pdet2-essick, Talbot2025, Heinzel:2025MonteCarlo}, as follows:
\begin{equation}
    \log p(\vec{d}|\vec{n}) =  \log p(\vec{d}|\vec{n}) = -\sum_\gamma n^\gamma \mu_{VT}^{\gamma} + \sum_i \log \biggl(\sum_\gamma n^\gamma \mu_{w}^{\gamma}(d_i)\biggr),
\end{equation}
\begin{equation}
    \sigma^2_{  \log p(\vec{d}|\vec{n})} = \sum_{\gamma}n_{\gamma}^2(\sigma_{VT}^{\gamma})^2+ \sum_{i}\frac{\sum_{\gamma}n_{\gamma}^2(\sigma_{w}^{\gamma}(d_i))^2}{\left(\sum_{\gamma}n_{\gamma}\mu_{w}^{\gamma}(d_i)\right)^2}.
\end{equation}
Here, the variance in the log-likelihood estimator $( \sigma^2_{  \log p(\vec{d}|\vec{n})})$ due to the finite number of Monte Carlo samples can become large for certain values of hyperparameters which should be rejected to avoid biases in the inferred population. We penalize the log-likelihood for hyperparameters that lead to $\sigma^2_{  \log p(\vec{d}|\vec{n})}$ larger than 1. To ensure finite gradients (as required by HMC) we impose this variance-cut asymptotically and modify the log-likelihood function as follows:
\begin{equation}
    \log \mathcal{P}(\vec{d}|\vec{n}) =  -\sum_\gamma n^\gamma \mu_{VT}^{\gamma} + \sum_i \log \biggl(\sum_\gamma n^\gamma \mu_{w}^{\gamma}(d_i)\biggr) - \log\{1+\left(\frac{1}{\sigma_{\log p(\vec{d}|\vec{n})}}\right)^{-n}\},
\end{equation}
where $n$ is a large positive integer which we choose to be 30~\citep{Callister:2023ParameterFree}. We have verified that all values of $n$ within the range $[30,75]$ yield comparable answers with very few hypersamples that correspond to large log-likelihood variance, ensuring unbiased population inference.

Given this likelihood, the GP prior on the logarithmic rate densities, and the priors on GP parameters, we construct our hyperposterior:
\begin{equation}
    \log p(\log \vec{n}, \mu_{GP}, \vec{\lambda}_{GP}, \sigma_{GP}| \vec{d}) = \log \mathcal{P}(\vec{d}|e^{\log\vec{n}}) -\frac{1}{2}C^{-1}_{\gamma,\gamma'}(\vec{\lambda}_{GP}, \sigma_{GP}) (\log n^{\gamma}-\mu_{GP})(\log n^{\gamma'}-\mu_{GP}) -\frac{1}{2}\log \rm{det}|C| +\pi(\mu_{GP}, \vec{\lambda}_{GP}, \sigma_{GP})
\end{equation}
where $\mu_{GP}$ is the mean function of the GP, $\vec{\lambda}_{GP}$ is the length scale, and $\sigma_{GP}$ of the covariance kernel $C$ which is constructed as follows:
\begin{equation}
    C_{\gamma,\gamma'} = \sigma^2\exp{\{-\frac{1}{2}\sum_{\theta\in(m_1,q,\chi_{eff},\chi_p)}\frac{(c_{\theta}^{\gamma}-c_{\theta}^{\gamma'})^2}{(\lambda_{GP}^{\theta})^2}\}}
\end{equation}
and, $\pi(\mu_{GP}, \vec{\lambda}_{GP}, \sigma_{GP})$ is the prior on GP hyperparameters. For $\sigma_{GP}$ we use a HalfNormal prior with scale=1, $\mu_{GP}$ a TruncatedNormal within the range (-15,15) with mean 0 and sigma=10, and LogNormal priors for each $\lambda_{GP}^{\theta}$ whose means and variances are chosen such that the smallest and largest separations between bin-centers~$(\{c_{\gamma}^{\theta}\})$ are within four sigma of the mean a priori. 
\label{sec:app-hyperpost}
\section{Scalability to Higher dimensions and Resolutions}
\label{sec:app-scale}
Given the pre-computability of $\mu_{w},\sigma_{w}, \mu_{VT},$ and $\sigma_{VT}$,~\citep{Sridhar:2025kvi, Ray:2023upk, Ray:2024hos} the dominant computational cost in evaluating the log-posterior is the inversion/factorization of the covariance matrix which has cubic time complexity with the number of bins/data points. Traditional implementations perform  Cholesky factorization of C and construct the posterior from whitened variables $\tilde{\vec{n}}$:

\begin{equation}
    \mathbf{C} = \mathbf{L}\mathbf{L}^{T}
\end{equation}
\begin{equation}
    \log\vec{n}-\mu_{GP}= \mathbf{L} \tilde{\vec{n}}
\end{equation}
in terms of which the posterior becomes:
\begin{equation}
    \log p(\tilde{\vec{n}},\mu_{GP}, \vec{\lambda}_{GP}, \sigma_{GP}| \vec{d}) =  \log \mathcal{P}(\vec{d}|e^{\mathbf{L} \tilde{\vec{n}}+\mu_{GP}}) -\frac{1}{2}(\sum_{\gamma}\tilde{n}^{\gamma})^2 + \log \pi(\mu_{GP}, \sigma_{GP}, \vec{\lambda}_{GP}).
\end{equation}
The dominant cost in evaluating this posterior comes from the Cholesky Factorization. Fortunately, due to the Latent Kronecker structure of the exponential quadratic Kernel:
\begin{eqnarray}
    \mathbf{C}&=&\bigotimes_{\theta}\mathbf{C}_{\theta}\\
    C_{\theta}^{\gamma\gamma'}&=& \sigma^{\frac{2}{n_{\theta}}}e^{-\frac{(c^{\gamma}_{\theta}-c_{\theta}^{\gamma'})^2}{2\lambda_{\theta}^2}},
\end{eqnarray}
and the commutivity of Cholesky factorization with the Kronecker product:
\begin{eqnarray}
    \mathbf{L}&=&\bigotimes_{\theta}\mathbf{L}_{\theta}\\
    \mathbf{C}_{\theta}&=& \mathbf{L}_{\theta}\mathbf{L}^{T}_{\theta},
\end{eqnarray}
this cost is additive over the number of BBH parameters: $O(\sum_{\theta}N_{bin,\theta}^{3})$. This enables tractable generalization of our GPU implementation to higher dimensions at the same resolution. However, the cubic time complexity in a single dimension remains a bottleneck for scalability to higher resolutions, and further exploration is necessary to circumvent this limitation.
\bibliography{sample701}{}
\bibliographystyle{aasjournalv7}



\end{document}